\documentclass[aps,pre,showpacs,showkeys,square,numbers,amssymb,amsmath,nobibnotes,nofootinbib,superscriptaddress]{revtex4}
\usepackage{docs}%
\usepackage{bm}%
\usepackage[colorlinks=true,linkcolor=red]{hyperref}
\usepackage{amsmath}
\usepackage{amsfonts}
\usepackage{amssymb}
\usepackage{eucal}
\usepackage{verbatim}
\usepackage{times}

\usepackage{graphicx}
\usepackage[sort&compress]{natbib}

\def\Tr{\mathrm{Tr}}
\begin{document}


\title{How many eigenvalues of a Gaussian random matrix are positive?}



\author{Satya N. Majumdar}
\author{C\'{e}line Nadal}
\affiliation{Laboratoire de Physique Th\'{e}orique et Mod\`{e}les
Statistiques (UMR 8626 du CNRS), Univ. Paris-Sud,
B\^{a}timent 100, 91405 Orsay Cedex, France}

\author{Antonello Scardicchio}
\affiliation{Abdus Salam International Centre for
Theoretical Physics, Strada Costiera 11, 34151 Trieste, Italy}
\affiliation{INFN, Sezione di Trieste, Strada Costiera 11, 34151 Trieste, Italy}

\author{Pierpaolo Vivo}
\affiliation{Abdus Salam International Centre for
Theoretical Physics, Strada Costiera 11, 34151 Trieste, Italy}


\date{\today}

\begin{abstract}
We study the probability distribution of the index ${\mathcal 
N}_+$, i.e., the number of positive eigenvalues of 
an $N\times N$ Gaussian random matrix. We show analytically
that, for large $N$ and large $\mathcal{N}_+$ with the fraction
$0\le c=\mathcal{N}_+/N\le 1$ of positive eigenvalues fixed, 
the index 
distribution 
$\mathcal{P}({\mathcal N}_+=cN,N)\sim\exp\left[-\beta N^2 \Phi(c)\right]$
where  
$\beta$ is the Dyson index characterizing the Gaussian ensemble.
The associated large deviation rate function  
$\Phi(c)$ is computed explicitly for all $0\leq c \leq 1$.
It is independent of $\beta$
and displays
a quadratic form modulated by a logarithmic singularity around 
$c=1/2$. 
As a consequence, the distribution of the index 
has a Gaussian form near the peak, but with a variance 
$\Delta(N)$ of index fluctuations growing as $\Delta(N)\sim \log N/\beta\pi^2$ 
for large $N$. For $\beta=2$, this result is independently confirmed 
against an exact finite $N$ formula, yielding $\Delta(N)= \log N/2\pi^2 +C+\mathcal{O}(N^{-1})$ for large $N$,
where the constant $C$ has the nontrivial value $C=(\gamma+1+3\log 2)/2\pi^2\simeq 0.185248...$ and $\gamma=0.5772...$ is the Euler constant. 
We also determine for large $N$ the probability that the interval 
$[\zeta_1,\zeta_2]$
is free of eigenvalues.
Part of these results have been announced in 
a recent letter [\textit{Phys. Rev. Lett.} {\bf 103}, 220603 (2009)].
\end{abstract}


\maketitle
\section{Introduction}

Statistical properties of eigenvalues of random matrices have been extensively 
studied for decades, stemming from the seminal work of Wigner~\cite{Wigner}. 
Random 
Matrix Theory (RMT) has successfully provided tools and methods to disparate 
areas of physics and mathematics~\cite{Mehta}, with countless 
applications so far. Statistics of several random variables
associated with random eigenvalues have been studied extensively.
This includes the length of a gap in the eigenvalue spectra, number
of eigenvalues in a given interval, the largest eigenvalue, the trace 
etc.~\cite{Mehta}. Most studies concerned with the probability
of {\em typical} fluctuation of such a random variable around its mean.

However, various recent applications of random matrix theory 
have posed questions regarding {\em atypical} large fluctuations of such 
random variables associated with the eigenvalues, thus triggering
a number of recent studies on the large deviation probabilities
of such random variables. This includes, for instance, the 
large deviation probability of the extreme (maximum and minimum) eigenvalues 
of Gaussian~\cite{DM,majumdar:060601,BEMN,Saito,ben} and
Wishart random matrices~\cite{vivo2007large,majumdar:060601,KC}, 
of the number of stationary points of
random Gaussian landscapes~\cite{BrayDean,fyod}, of the distribution
of free energies in mean-field spin glass models~\cite{PR,MG},
of the conductance and shot 
noise power in chaotic mesoscopic 
cavities~\cite{vivo:216809,conductance},
of the entanglement entropy of a pure random state of a bipartite
quantum system~\cite{facchi:050502,NMV,pasquale,pp2010} and of the mutual
information in multiple input multiple output (MIMO)
channels~\cite{kaz}. In addition, random matrix theory has been used to
understand large deviation properties of various observables
in the so called vicious walker (or nonintersecting Brownian motion) 
problem~\cite{Schehr1,NM,Schehr2,FMS}. The purpose of the present paper
is to provide a detailed analysis of the large deviation properties
of another natural random variable for large Gaussian matrices, namely
the fraction $c$ of positive eigenvalues of an    
$N\times N$ Gaussian matrix. Part of the main results
presented here were announced in a recent Letter~\cite{letter}. 
We will explain shortly why this fraction $c$ is a natural observable that 
arises in
a number of physical situations. But before we do that, it is useful
to recall some well-known facts about Gaussian matrices. 

There are three families of Gaussian random matrices with real spectrum: orthogonal
(GOE), unitary (GUE) and symplectic (GSE). The $N\times N$ matrices belonging to these families are real symmetric, complex hermitian and quaternion self-dual respectively,
whose entries are independent Gaussian variables (real, complex or quaternions) labeled by
the Dyson index $\beta=1,2,4$ respectively. 
The probability distribution of the entries of a matrix $\mathbf{M}$  
is then given by the Gaussian weight:
\begin{equation}
\mathcal{P}(\mathbf{M})\propto\exp\left(-\frac{\beta}{2}(\mathbf{M},\mathbf{M})\right)
\label{entrydist}
\end{equation}
where $(\mathbf{M},\mathbf{M})$ stands for the inner product on the space of matrices invariant under orthogonal, unitary and symplectic transformation respectively. Explicitly, one has:
\begin{align}
(\mathbf{M},\mathbf{M}) &=\Tr(M^2), &\quad \beta=1 &\quad \mathrm{GOE}\\
(\mathbf{M},\mathbf{M}) &=\Tr(M^\star M), &\quad \beta=2 &\quad \mathrm{GUE}\\
(\mathbf{M},\mathbf{M})&=\Tr(M^\dagger M), &\quad \beta=4 &\quad \mathrm{GSE}
\end{align}
where $\star$ denotes hermitian conjugation and $\dagger$ the quaternion self-dual.
The celebrated result by Wigner states that for large matrix size $N$, 
the average density of eigenvalues (all real) for such ensembles 
has a $\beta$-independent semicircular form~\cite{Wigner,Mehta}
\begin{equation}\label{sc}
\rho_{\rm{sc}}(\lambda,N)=\sqrt{\frac{2}{N\pi^2}}\left[1-\frac{\lambda^2}{2N}\right]^{1/2}
\end{equation} 
which vanishes identically at the two edges $\pm\sqrt{2N}$ and is normalized 
to unity.
Clearly, the mean spacing between eigenvalues in the bulk, i.e., close to the 
origin, behaves for large $N$ as $\delta_N=1/{\left( N\rho_{\rm 
sc}(0)\right)}=\pi/\sqrt{2N}$.
 
A natural and much studied question that goes back to Dyson~\cite{Dyson} 
is: how many eigenvalues are there in a
given interval $[a,b]$ on the real line? 
Clearly this number $\mathcal{N}_{[a,b]}$ is a random variable that fluctuates 
from 
one sample to another. Its mean value, for large $N$, is easy to 
compute by integrating the semi-circular average density in \eqref{sc} over
the interval $[a,b]$: $\langle \mathcal{N}_{[a,b]}\rangle =N \int_a^b \rho_{\rm 
{sc}}(\lambda,N) d\lambda$. But how does this number fluctuate from
one sample to another? Dyson studied this number fluctuation in the
so called bulk limit, i.e., he focused on a small symmetric interval
around the origin $[-\delta_N L/2, \delta_N L/2]$ where 
$\delta_N=\pi/\sqrt{2N}$
is the mean bulk spacing and $L$ is kept fixed while one takes the $N\to 
\infty$ limit. Let $\mathcal{N}_L$ denote the number of eigenvalues in this
interval. Clearly, the mean number of eigenvalues $\langle 
\mathcal{N}_L\rangle = L$.
But Dyson also computed the variance of $\mathcal{N}_L$ in the large $N$ 
limit (with
$L$ fixed) and showed that for large $L$ the variance grows logarithmically
with $L$ 
\begin{equation}
\langle (\mathcal{N}_L-L)^2\rangle \approx \frac{2}{\pi^2 \beta} \log (L) + 
B_\beta  
\label{vardyson}
\end{equation}
and the constant $B_\beta$ was computed by Dyson and 
Mehta~\cite{Dyson-Mehta}. Thus the {\em typical} fluctuations of 
$\mathcal {N}_L$ grow as $\sqrt{\log L}$ for large $L$. More recently, even 
the higher 
moments of $\mathcal{N}_L$ (in the $N\to \infty$ limit with $L$ fixed) were 
computed
which proved that on a scale of $\sqrt{\log L}$ around the mean $L$, the
random variable $\mathcal{N}_L$ has a Gaussian distribution~\cite{CL,FS}. 

Here our focus will be on a different limit, namely we study
the statistics of the number of eigenvalues, not on a small 
symmetric interval around the origin (i.e, the bulk limit), but rather
on the full unbounded interval $[0,\infty]$. In other words, we
are interested simply in the distribution of the number of positive
eigenvalues $\mathcal{N}_{+}$ (called the index) of a Gaussian random matrix 
$\mathbf{M}$.  
Since the average density of states is symmetric in $\lambda$, it 
is clear that on average there are $\langle 
\mathcal{N}_{+}\rangle=N/2 $ positive eigenvalues.  
Clearly the index $\mathcal{N}_{+}$ fluctuates from
one realization of the matrix to another and in this paper,
we are precisely interested in the fluctuation properties
of the random variable $\mathcal{N}_{+}$, i.e., in the
full probability distribution ${\mathcal P}(\mathcal{N}_{+},N)$.
Evidently, $0\le \mathcal{N}_{+} \le N$. Also, the number of negative
eigenvalues $\mathcal{N}_{-}=N-\mathcal{N}_{+}$ is distributed
identically as the number of positive eigenvalues $\mathcal{N}_{+}$
by virtue of the Gaussian symmetry, indicating
${\mathcal P}(\mathcal{N}_{+},N)={\mathcal
P}(N-\mathcal{N}_{+},N)$. Hence the distribution ${\mathcal 
P}(\mathcal{N}_{+},N)$ of $\mathcal{N}_{+}$ is clearly symmetric around its 
mean value 
$\langle \mathcal{N}_{+}\rangle =N/2$. It thus suffices to study the range
$N/2\le \mathcal{N}_{+} \le N$.
 
So, why are we interested in this index distribution?
This question naturally arises in the study of the stability
patterns associated with a multidimensional potential
landscape $V(x_1,x_2,\ldots,x_N)$~\cite{Wales}. 
For instance, in the context of glassy systems,
the point $\{x_i\}$ represents a configuration of the system
and $V(\{x_i\})$ is just the energy of the configuration~\cite{CGG}.
Similarly, in the context of disordered systems or spin glasses, 
$V(\{x_i\})$ may represent the free energy landscape. 
In the context of string theory, $V$ may represent the potential
associated with a moduli space~\cite{Douglas}. 
Typically such an 
$N$-dimensional 
landscape has many stationary points (minima, maxima and saddles)
with complex stability patterns that play an important role both in statics
and dynamics of such systems~\cite{Wales}. The stability of a stationary point
of this $N$-dimensional landscape is decided by the $N$ real eigenvalues of
the $(N\times N)$ Hessian matrix $M_{i,j}= \left[\partial^2 V/{\partial x_i 
\partial 
x_j}\right]$ which is symmetric. If all the eigenvalues are positive 
(negative), the stationary point is a local minimum (local maximum).
If some, but not all, are positive then the stationary point is
a saddle. The number of positive eigenvalues (the index), $0\le  {\mathcal 
N_{+} }\le N$, is then a key object that determines in how many directions
the stationary point is stable. Given a  
random potential $V$, the entries of the 
Hessian
matrix at a stationary point are usually correlated. However, in many
situations, important insights can be obtained by ignoring these
correlations and just assuming the entries of the Hessian matrix are
just independent Gaussian variables. This then
leads to the study of the statistics of index for a GOE matrix.
This toy model, called the 
random 
Hessian model (RHM), has been studied extensively in the context of disordered
systems~\cite{CGG}, landscape based string 
theory~\cite{aazami} and 
also in quantum cosmology~\cite{Mersini-Houghton05}.
Although in RHM $\beta=1$, it is quite natural to study the index distribution 
for other Gaussian ensembles, namely for GUE ($\beta=2$) and GSE ($\beta=4$).

For the GOE ($\beta=1$), the statistics of $\mathcal{N}_+$ was studied by 
Cavagna {\em et al.}~\cite{CGG}
using supersymmetric replica methods and some additional approximations. 
They argued that around its mean value $N/2$, the random variable
$\mathcal{N}_+$ has {\em typical} fluctuations of $\mathcal{O}(\sqrt{\log N})$ for large 
$N$. Moreover, the distribution of these typical fluctuations 
is Gaussian.
In other words, over a region of width $\sqrt{\log N}$, the distribution
for large $N$
is given by~\cite{CGG}
\begin{equation}
{\mathcal P}(\mathcal{N}_+,N) \approx \exp\left[- \frac{\pi^2}{2\log
(N)}\left(\mathcal{N}_+-N/2\right)^2\right]     
\label{gauss1}
\end{equation}
implying that for $\beta=1$, $\Delta(N)=\langle (\mathcal{N}_{+}-N/2)^2\rangle \approx 
\log(N)/{
\pi^2}$ for large $N$.

On the other hand, this Gaussian form does not describe the {\em atypically}
large fluctuations of $\mathcal{N}_{+}$. For example, in the extreme limit
when $\mathcal{N}_{+}=N$, the probability that all eigenvalues are positive
${\mathcal P}(\mathcal{N}_+=N,N)$ was computed recently for large $N$
and for all $\beta$~\cite{DM},
\begin{equation}
{\mathcal P}(\mathcal{N}_+=N,N) \approx \exp\left[-\beta \theta N^2 \right]; 
\quad
\theta=\frac{1}{4}\log(3).
\label{dm1}
\end{equation}
This question of the probability of extreme large fluctuation of 
$\mathcal{N}_+$ (fluctuation on a scale $\sim \mathcal{O}(N)$ around its mean $N/2$)
naturally came up in several recent contexts
such as in landscape based string
theory~\cite{aazami}, quantum cosmology~\cite{Mersini-Houghton05} and in the distribution
of the number of minima of a random polynomial \cite{randpol}.

These two rather different forms of the distribution ${\mathcal 
P}(\mathcal{N}_+,N)$ in the two limits, namely in the vicinity
of $\mathcal{N}_+=N/2$ (over a scale of $\sqrt{\log N}$) (as in 
\eqref{gauss1}) and when $\mathcal{N}_+=N$ (as in \eqref{dm1})
raise an interesting question: what is the form of the distribution
${\mathcal P}(\mathcal{N}_+,N)$ for intermediate values of $N/2<< 
\mathcal{N}_{+}<N$?
In other words, how does one interpolate between the limits of {\em 
typically} small
and {\em atypically} large fluctuations?
To answer this question, it is natural to set $\mathcal{N}_{+}=cN$ where 
the intensive variable $0\le 
c\le 1$ denotes the fraction of positive eigenvalues and 
study the large $N$ limit of the distribution ${\mathcal
P}(cN,N)$ with $c$ fixed. Again, due to the Gaussian symmetry, 
${\mathcal P}(cN,N)={\mathcal
P}((1-c)N,N)$ and it is sufficient to restrict $c$ in the range $1/2\le c\le 
1$.

In a recent 
Letter \cite{letter}, we computed the large $N$ limit of 
the distribution $\mathcal{P}(cN,N)$ in the full range $0\le c\le 1$
for all $\beta>0$ and showed that
\begin{equation}
{\mathcal P}(cN,N) \approx \exp\left[-\beta\, N^2\, \Phi(c)\right]
\label{ldv0}
\end{equation}
where the rate function $\Phi(c)=\Phi(1-c)$, independent of $\beta$, was 
computed explicitly for 
all $1/2\le c\le 1$\footnote{Hereafter, the notation $\approx$ stands for
the precise asymptotic law $\lim_{N\to\infty} 
\frac{-\log \mathcal{P}(cN,N)}{\beta N^2}=\Phi(c)$.}.
The fact that the logarithm of the probability is $\sim \mathcal{O}(N^2)$ for fixed $c$
is quite natural, as it represents the free energy of
an associated Coulomb fluid of $N$ charges (eigenvalues) (to be discussed
in detail later).
The Coulomb energy of $N$ charges clearly scales as $\sim \mathcal{O}(N^2)$.
In the
limit $c\to 1$, we get $\Phi(1)=\theta= \log(3)/4$ in agreement with
(\ref{dm1}). The distribution is thus highly non-Gaussian near
its tails. In the opposite
limit $c\to 1/2$, we find a marginally quadratic
behavior, modulated by a logarithmic singularity
\begin{equation}
\label{gausslog1}
\Phi(c)\simeq -\frac{\pi^2}{2}\frac{(c-1/2)^2}{\log(c-1/2)}.
\end{equation}
Setting $c=\mathcal{N}_+/N$ and substituting this form in
\eqref{ldv0}, we find that in the vicinity of $\mathcal{N}_{+}=N/2$
and over a scale of $\sqrt{\log N}$, indeed one recovers the
Gaussian distribution 
\begin{equation}
{\mathcal P}(\mathcal{N}_+,N) \approx \exp\left[- 
\frac{\beta\, \pi^2}{2\log
(N)}\left(\mathcal{N}_+-N/2\right)^2\right]
\label{gauss10}
\end{equation}
thus proving that the variance $\Delta(N)=\langle (\mathcal{N}_{+}-N/2)^2\rangle
\approx \log(N)/{\beta \pi^2}$ for large $N$ and for all $\beta$.
For $\beta=1$, this perfectly agrees with the results of Cavagna 
{\em et al.}~\cite{CGG}.

In addition to obtaining the full distribution ${\mathcal P}(cN,N)$ 
of the fraction of positive eigenvalues $c$, our Coulomb gas approach also
provides a new method of finding solutions to singular integral
equation with two disconnected supports, as discussed in detail
later. This method 
is rather general
and can be fruitfully applied to other related problems in RMT, an
example is later provided in the paper in calculating
the probability that an interval $[\zeta_1,\zeta_2]$ is free
of eigenvalues, i.e., there is a gap $[\zeta_1,\zeta_2]$ in the
spectrum.
The details of these calculations are somewhat involved and
were not presented in our previous Letter~\cite{letter}. The purpose of this
paper is to provide these details which we believe will be important
for other problems as well.

The paper is organized as follows. In Section II.A we set up the 
problem and show that the rate function can be computed via the solution of a 
singular integral equation on a disconnected support. In  
subsections II.B and II.C, 
we provide two different strategies to find such a solution, the first based on 
a scalar Riemann-Hilbert ansatz and the second based on an iterated 
application of a theorem by Tricomi. In subsection II.D we derive the free 
energy of the 
associated Coulomb gas and the large deviation function $\Phi(c)$ associated
with the index distribution. In subsection II.E we provide an asymptotic
analysis of $\Phi(c)$ near $c=1/2$ and  
determine the variance 
of the index for large matrix size $N$. 
In section III we provide details of numerical simulations.
As an application
of the general method for solving two-support integral equation, we compute in 
section IV, 
the 
probability that a Gaussian random matrix has a gap $[\zeta_1,\zeta_2]$ in the 
spectrum.
In section V we offer 
a derivation of a determinantal 
formula for the variance of the index at finite $N$ for $\beta=2$.
Finally, we conclude with a summary in section VI.

\section{The probability distribution of the index}
\label{prob-section}

\subsection{Setting and Notation}

We consider the standard Gaussian ensembles of random matrices with Dyson 
index $\beta=1,2,4$, corresponding to real, complex and quaternion 
entries respectively.
The probability distribution of the entries 
is given in \eqref{entrydist} and consequently the
joint probability density of eigenvalues reads~\cite{Mehta}
\begin{equation}\label{eq:jpdfEV}
\mathcal{P}(\lambda_1,\ldots,\lambda_N)=\frac{1}{Z_N}e^{-\frac{\beta}{2}\sum_{i=1}^N\lambda_i^2}\prod_{j<k}|\lambda_j-\lambda_k|^\beta
\end{equation}
where $Z_N$ is the normalization constant which can be explicitly computed via 
a Selberg-like integral~\cite{Mehta} and to leading order for large $N$, 
$Z_N \approx \exp(-\beta\Omega_0 N^2)$
where $\Omega_0=(3+2\log 2)/8$~\cite{DM}. 

We wish to compute the probability distribution $\mathcal{P}(\mathcal{N}_{+},N)$ of the index $\mathcal{N}_{+}$, defined as the number of positive eigenvalues of the 
$N\times N$ matrix $\mathbf{M}$:
\begin{equation} \label{eq:NplusDef}
\mathcal{N}_{+}=\sum_{i=1}^N \theta(\lambda_i)
\end{equation}

By definition:
\begin{equation}
\label{probdef}
\mathcal{P}(\mathcal{N}_{+},N)=
\frac{1}{Z_N}\int_{(-\infty,\infty)^N}
\prod_i d\lambda_i e^{-\frac{\beta}{2}
\sum_{i=1}^N\lambda_i^2}\prod_{j<k}|\lambda_j-\lambda_k|^\beta
\delta\left(\mathcal{N}_{+}-\sum_{i=1}^N \theta(\lambda_i)\right)
\end{equation}
We will set $\mathcal{N}_{+}=cN$ where $0\le c\le 1$ is the fraction
of positive eigenvalues. As mentioned in the introduction, due to the
Gaussian symmetry, the number of positive eigenvalues $\mathcal{N}_{+}$ will 
have the
same distribution as the number of negative eigenvalues 
$\mathcal{N}_{-}=N-\mathcal{N}_{+}$. Hence, 
$\mathcal{P}(cN,N)=\mathcal{P}\left((1-c)N,N\right)$ (the distribution
is symmetric around $c=1/2$).
Thus, it is sufficient to focus only on the range $1/2\le c\le 1$.

The evaluation of the $N$-fold integral \eqref{probdef} in the large $N$ limit consists of the following steps: first, we write the integrand (ignoring the delta function) as, $\exp\left[-\beta E(\{\lambda_i\})\right]$
with $E(\{\lambda_i\})= -(1/2) \sum_{j\ne k} \log |\lambda_j-\lambda_k| +(1/2)\sum_i \lambda_i^2$. 
Written in this form, the integral has a natural
interpretation as the partition function of a Coulomb gas in equilibrium at inverse temperature $\beta$. We can identify $\lambda_i$'s as the
coordinates of the charges of a $2$-d fluid confined on
the real axis. The charges repel each other
via the $2$-d logarithmic Coulomb potential and are confined
by a quadratic external potential. 
Then $E$ is the energy of this Coulomb gas. Furthermore, the Coulomb energy
scales, for large $N$, as $\sim \mathcal{O}(N^2)$ (since it involves pairwise
interaction between $N$ charges). In contrast, the external potential
energy scales as $\sim \lambda_{\rm typ}^2 N$ where $\lambda_{\rm typ}$ is
a typical eigenvalue. 
Balancing the two energy scales, one finds that a typical
eigenvalue scales as $\lambda_{\rm typ}\sim \sqrt{N}$ for large $N$.

The next step is to evaluate this partition function
of the Coulomb gas in the large $N$ limit via the saddle point method.
In the large N limit, the eigenvalues become rather dense and
one can then take a continuum limit where one replaces the
integration over the discrete eigenvalues by a functional
integral over the density of these eigenvalues.
Originally introduced by Dyson~\cite{Dyson}, this procedure (see
also ~\cite{Forrester}) 
has recently been successfully used in a number of different contexts.
These include the computation of  
the extreme eigenvalue 
distribution of
Gaussian~\cite{DM,majumdar:060601} and 
Wishart random matrices~\cite{vivo2007large,majumdar:060601,KC}, 
counting the number of stationary points of
random Gaussian landscapes~\cite{BrayDean,fyod}, and computing
the distribution of the  
bipartite quantum entanglement ~\cite{facchi:050502,NMV,pasquale}.
In addition, this method has also been used recently in systems
such as  
nonintersecting
fluctuating interfaces in presence of a substrate~\cite{NM},
in computing the distribution of conductance and shot noise power in mesoscopic 
cavities~\cite{vivo:216809,conductance}
and in the study of multiple input multiple output (MIMO) 
channels~\cite{kaz}.
   
Dyson's prescription requires first a coarse-graining procedure, 
where one sums over (partial 
tracing) all microscopic
configurations of $\lambda_i$'s compatible with a 
fixed charge density function $\varrho_N(\lambda)=N^{-1}\sum_i
\delta(\lambda-\lambda_i)$. Secondly, one performs a functional integral
over all possible positive charge densities $\varrho_N(\lambda)$ normalized to unity.
Finally the functional integral is carried out in the large $N$ limit by the 
saddle point method. 

Following this prescription, we introduce a continuum fluid representation 
for the Coulomb cloud of eigenvalues with density 
$\varrho_N(\lambda)=N^{-1}\sum_{i=1}^N\delta(\lambda-\lambda_i)$.
Since $\lambda_{\rm typ} \sim \sqrt{N}$, it follows that the normalized
density should have the scaling form 
$\varrho_N(\lambda)=N^{-1/2}f_{c}(\lambda/\sqrt{N})$ for large $N$.
The scaled density $f_{c}(x)$ satisfies the obvious normalization conditions:
\begin{align}
\label{n1}\int_{-\infty}^\infty dx f_{c}(x) &=1\\
\label{n2}\int_{-\infty}^\infty dx \theta(x) f_{c}(x) &=c
\end{align}
where we have set $\mathcal{N}_{+}=cN$ with $1/2\le c\le 1$ being
the fraction of positive eigenvalues.
The probability density \eqref{probdef} can then be rewritten as a functional 
integral over $f_{c}(x)$ as:
\begin{equation}
\label{prob}
\mathcal{P}(\mathcal{N}_{+}=cN,N)=\frac{Z_c(N)}{Z_N}
\end{equation}
where the numerator $Z_c(N)$ reads:
\begin{align}
\nonumber Z_c(N) &=\int\mathcal{D}[f_{c}(x)]\exp\left\{-\frac{\beta}{2}N^2\left[\int_{-\infty}^\infty dx x^2 f_{c}(x)-\int_{-\infty}^\infty\int_{-\infty}^\infty 
dx dx^\prime f_{c}(x) f_{c}(x^\prime)\log|x-x^\prime|+\right. \right. \\
& \left. \left. +A_1\left(\int_{-\infty}^\infty dx\theta(x)f_c(x)-c\right)+ A_2\left(\int_{-\infty}^\infty dx f_c(x)-1\right) \right] \right\}\label{ZCN}
\end{align}
where $A_1,A_2$ are Lagrange multipliers enforcing the normalization conditions \eqref{n1} and \eqref{n2}.

We define the action $S[f_c(x)]$ as:
\begin{equation}\label{action}
S[f_c(x)]=
\int_{-\infty}^\infty dx x^2 f_{c}(x)-\int_{-\infty}^\infty\int_{-\infty}^\infty 
dx dx^\prime f_{c}(x) f_{c}(x^\prime)\log|x-x^\prime|+
A_1\left(\int_{-\infty}^\infty dx\theta(x)f_c(x)-c\right)+ 
A_2\left(\int_{-\infty}^\infty dx f_c(x)-1\right). 
\end{equation}
Evaluating \eqref{ZCN} by the 
method of steepest descent and using the large $N$ asymptotics of the 
denominator $Z_N$ in \eqref{prob} gives,
to leading order for large $N$,
\begin{align}
\label{znum} Z_c(N) &\approx \exp\left(-\frac{\beta}{2}N^2 
S[f_c^\star(x)]\right)\\
\label{zden} Z_N &\approx \exp(-\beta\Omega_0 N^2)
\end{align}
where $\Omega_0=(3+2\log 2)/8$ \cite{DM} and $f_c^\star(x)$ is the solution of
the saddle point equation
\begin{equation}\label{saddle point}
0=\frac{\delta S[f_c(x)]}{\delta f_c}=x^2+A_1\theta(x)+A_2-
2\int_{-\infty}^\infty dx^\prime f_c^\star(x^\prime)\log|x-x^\prime|
\end{equation} 
The function $f_c^\star(x)$ can be interpreted as the equilibrium (or 
optimal) 
charge density of the eigenvalue fluid, given a fixed fraction $c$ of
positive charges. Once we obtain the solution $f_c^\star(x)$ of the integral
equation \eqref{saddle point}, we can evaluate the saddle point action
in \eqref{znum}, and together with \eqref{zden} one then gets
the index distribution
\begin{equation}\label{largedevlaw}
\mathcal{P}(cN,N)=\frac{Z_c(N)}{Z_N}\approx\exp\left(-\beta N^2 \underbrace{\left[\frac{1}{2}S[f_c^\star(x)]-\Omega_0\right]}_{\Phi(c)}\right)
\end{equation}
where $\Phi(c)$ is the large deviation function.

Thus all we have to do is to solve the saddle point equation \eqref{saddle 
point}
for a fixed $1/2\le c\le 1$. To avoid the Lagrange multipliers, it is
convenient to differentiate \eqref{saddle point} with respect to $x$ 
and for ($x\ne 0$), one gets the integral equation
\begin{equation}\label{tricomi}
x = \mathrm{Pr}\int_{-\infty}^\infty\frac{f_c^\star(x^\prime)}
{x-x^\prime}dx^\prime
\end{equation}
(where $\mathrm{Pr}$ denotes Cauchy's principal value), supplemented with the constraints:
\begin{align}
\label{n1bis}\int_{-\infty}^\infty dx f_{c}^\star(x) &=1\\
\label{n2bis}\int_{0}^\infty dx f_{c}^\star(x) &=c
\end{align}
Singular integral equations of this type have been studied by 
Tricomi~\cite{Tricomi}, 
who derived an explicit formula for the solution $f_c^\star(x)$ in the case
when the solution is nonzero over  
a {\em single} finite connected support $x\in [L_1,L_2]$ where $L_1$ and 
$L_2$ are
respectively the lower and the upper end of the support.  
Tricomi's theorem states that
the general solution $f(x)$ to singular integral equations of the form
\begin{equation}
\label{intequation1}
g(x)=\mathrm{Pr}\int_{L_1}^{L_2}\frac{f(x^\prime)}{x-x^\prime}dx^\prime
\end{equation}
over the interval $[L_1,L_2]$ with 
$L_1<L_2$ (where the source function $g(x)$ is given and arbitrary) 
is~\cite{Tricomi}:
\begin{equation}
\label{formulatricomi}
f(x)=-\frac{1}{\pi^2\sqrt{(L_2-x)(x-L_1)}}
\left[\mathrm{Pr}\int_{L_1}^{L_2}\frac{\sqrt{(L_2-x^\prime)(x^\prime-L_1)}}
{x-x^\prime}g(x^\prime)dx^\prime+B_1\right]
\end{equation}
where $B_1=-\pi \int_{L_1}^{L_2} f(x)dx$ is a constant.

Let us then first assume that indeed the solution $f_c^\star(x)$ of 
\eqref{tricomi}, with the source function $g(x)=x$, has a {\em single}
support over $[L_1,L_2]$. Substituting $g(x)=x$, one can evaluate 
the integral in \eqref{formulatricomi} explicitly to obtain
\begin{equation}
f_c^\star(x)= \frac{1}{8\pi 
\sqrt{(L_2-x)(x-L_1)}}\left[(L_2-L_1)^2+4(L_2+L_1)x-8x^2+8\right]
\label{explicit}
\end{equation}
where we have used the normalization condition 
$\int_{L_1}^{L_2}f_c^\star(x)dx=1$
to set the constant $B_1=-\pi$.
There are
two unknown constants $L_1$, $L_2$ which are to be fixed
from the constraint \eqref{n2bis} and
the consistency condition that the solution $f_c^\star(x)$ (which represents
a density) must be non-negative over $[L_1,L_2]$. At the two endpoints $L_1$ 
and 
$L_2$, the solution either vanishes or has an inverse square root divergence 
(which is integrable). If we try to evaluate these constants, it is easy to 
check that a non-negative consistent solution is possible only for
two limiting values of $c$, namely $c=1/2$ and $c=1$. Let us
discuss these two cases first.
\vskip 0.2cm

\noindent {\bf {The case $c=1/2$:}} In this case, the solution must be 
symmetric which indicates $L_1=-L_2$. In addition, it is clear physically
that the solution must vanish at the endpoints $L_1$ and $L_2$. This
fixes $L_2=-L_1=\sqrt{2}$ and the solution
in \eqref{explicit} reduces 
to the Wigner semicircle law, 
namely
\begin{equation}
f_{1/2}^\star(x)= \frac{1}{\pi}\,\sqrt{2-x^2}.
\label{wigner1}
\end{equation}
This is reassuring and
is expected
for the following reason: if there was no constraint at all on the fraction of positive 
eigenvalues, the system would naturally choose to have half the eigenvalues 
positive and half negative on average, implying $\langle {\mathcal N}_{+}\rangle=N/2$, and the 
equilibrium charge density would be the standard Wigner's semicircle law. 

\vskip 0.2cm

\noindent {\bf {The case $c=1$:}} In the other extreme limit $c=1$ where 
all the eigenvalues are forced to be positive, one can again find a
consistent solution from \eqref{explicit} that satisfies all
the constraints and is given by
\begin{equation}
f_1^\star(x)= \frac{1}{2\pi}\,\sqrt{\frac{L-x}{x}}\, \left[L+2x\right]
\label{DM1}
\end{equation}
where $L=2\sqrt{2/3}$. In this case, the support is over $[0,L]$
with $L_1=0$, $L_2=L$. Note that this solution vanishes at
the upper edge $x=L$ and diverges as $x^{-1/2}$ at the lower edge $x=0$.
This explicit solution was first obtained in ~\cite{DM}.

\begin{figure}[htbp]
\includegraphics[height=6cm]{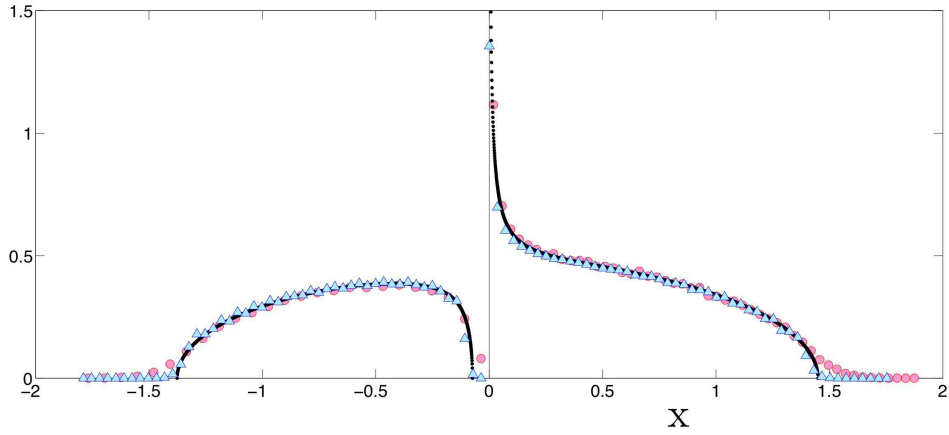}
\caption{Analytical density $f_c^\star(x)$ in \eqref{solutionforf} for $c=0.6$
(solid black) together with results from i) (red) numerical diagonalization
of
$10^6$ matrices of size $20\times 20$, where only
samples having $12$ positive eigenvalues were retained
for the statistics ($c=0.6$), and ii) (blue) Montecarlo simulations of
the Coulomb fluid with $N=50$ particles.}
\label{fig:densnum}
\end{figure}

It turns out that for other values of $1/2<c<1$, there is no {\em 
single support} 
solution \eqref{explicit} that satisfies the constraint \eqref{n2bis}
and is non-negative for all $x\in [L_1,L_2]$. To see what is going wrong, it
was instructive to perform numerical simulation (the details of which
will be described later) for $1/2<c<1$. 
For example, for $c=0.6$, the optimal density is given in Fig. 
(\ref{fig:densnum}). It is evident from the figure that for $c=0.6$,
indeed there are two disconnected supports of the optimal charge
density $f_c^\star(x)$. 

\begin{figure}[htbp]
\includegraphics[bb = 0 0 240 157, height=6cm]{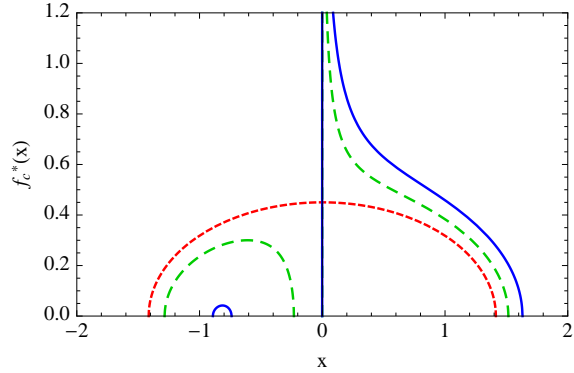}
\caption{The optimal density of eigenvalues $f_c^\star(x)$ (Eq. 
\eqref{solutionforf})
for $c=1/2$ (red),
$3/4$ (green)  and $0.995$ (blue).}
\label{fig:aLrho}
\end{figure}

A similar feature actually holds for all
$1/2<c<1$. As $c\to 1$ from below, the area under the left support
vanishes and we are left with a single support over $[0,L]$ as
in \eqref{DM1}. On the other hand, as $c$ decreases continuously,
the area under the left support grows and the upper edge of the
left support (always on the negative side) also increases. Finally
when $c$ hits $1/2$, the two supports merge into a single support, 
symmetric
about the origin, and reduces to the Wigner semicircle law (see Fig. 
(\ref{fig:aLrho})).

Hence, it is not surprising that we cannot obtain any consistent {\em single support}
solution using Tricomi's result in \eqref{explicit} for $1/2<c<1$, as
the optimal density does not have a single support but rather two 
disconnected supports.
The technical reason for the two-support solution can indeed be traced back to 
the jump discontinuity at $x=0$ due to the Heaviside theta function
in the saddle point equation \eqref{saddle point}. 
So, the main technical challenge is how to obtain analytically an explicit 
two-support solution of the integral equation \eqref{tricomi}
for all $1/2<c<1$, given that we cannot use the Tricomi solution any more.
This is an interesting
mathematical challenge since such two-support solutions appear
in other problems as well and a general method would be very useful.  
This is what we achieve here as detailed 
in the next two subsections.
In fact we will present two different approaches producing
the same results. But before we get into the technical details of the 
two methods, it may be useful
to summarize here the main result. 

We show that the solution of \eqref{tricomi} satisfying the constraints 
\eqref{n1bis} and \eqref{n2bis} and the condition of non-negativity, for all 
$1/2\le c\le 1$ is given by
\begin{equation}\label{solutionforf} 
f_c^\star(x)=\frac{1}{\pi}\sqrt{\frac{L(a)-x}{ax}}\sqrt{(ax+L(a))(x+(1-1/a)L(a))} 
\end{equation} 
where 
\begin{equation} 
L(a)=\frac{a\sqrt{2}}{\sqrt{a^2-a+1}} 
\label{La}
\end{equation} 
and the parameter $a$ is determined implicitly as a function of $c$ from 
\eqref{n2bis} by the condition: 
\begin{equation}\label{normcond} \int_0^1 dy 
\sqrt{\frac{1-y}{y}}\sqrt{y^2+y+\frac{a-1}{a^2}}=
\frac{\pi}{2}\left(1-\frac{a-1}{a^2}\right)c. 
\end{equation} 
For general $c$, the equilibrium density \eqref{solutionforf} 
has support on the union of two disconnected intervals 
\begin{equation}\label{interval}
[-L(a)/a,-(1-1/a)L(a)]\cup (0,L(a)].
\end{equation}
One can easily check that in the two limiting cases $c=1/2$ and $c=1$,
our general solution reduces respectively to \eqref{wigner1}
and \eqref{DM1}.

\begin{itemize}

\item $c=1$: this corresponds to having no negative eigenvalues at all, thus 
the equilibrium density must match the solution in \cite{DM} at $z=0$. This 
is achieved as long as $a\to 2$ and thus $L(a)\to \sqrt{8/3}$ as expected 
(compare to \cite{DM}). Then the blob of negative eigenvalues in 
\eqref{solutionforf} (see \eqref{interval}) collapses to a single point and 
vanishes. 
\item $c=1/2$: this case represents the usual Wigner's semicircle 
and is recovered from \eqref{solutionforf} when $a\to 1$ and consequently 
$L(a)\to\sqrt{2}$. In this case, the support \eqref{interval} becomes compact 
$[-\sqrt{2},\sqrt{2}]$ as it should. 

\end{itemize}

In the next two subsections, we provide two 
alternative derivations of \eqref{solutionforf}, the first one based on a 
scalar Riemann-Hilbert ansatz and the second one based on an iterated 
application of Tricomi's single support solution.

\subsection{Method I: proof of \eqref{solutionforf} via Riemann-Hilbert ansatz}

In the context of counting of planar diagrams, Brezin et. al.~\cite{BIPZ}
encountered singular integral equation of the type \eqref{intequation1}
with a single support solution. They did not use the explicit Tricomi 
solution, but instead developed an alternative method using
a scalar Riemann-Hilbert ansatz. This method
makes use of properties of analytic functions in the complex plane.
Even though the method requires 
making a guess or ansatz (verified a posteriori), it turns out to be 
rather useful. This method can be generalized in a straightforward manner
to the case when the solution has multiple disconnected supports
and has been used before in other contexts [an example in a 
specific case can be found in the appendix of ~\cite{conductance}, see also \cite{akemann}]. Let us 
illustrate below the main idea behind this method.

Let us consider the singular integral equation
\begin{equation}\label{prpr}
g(x)=\mathrm{Pr} \int_{\mathcal{S}} 
dx^\prime \frac{f(x^\prime)}{x-x^\prime}
\end{equation} 
where the solution $f(x)$ has support on the union of a finite number of 
intervals $\mathcal{S}=\bigcup_{k=1}^M [\alpha_k,\beta_k]$ on the real line
and is normalized to unity: $\int_{-\infty}^{\infty} f(x)dx=1$. The next step
is to define a complex function $F(z)$ (without the principal part)
\begin{equation}\label{Frho}
F(z)=\int_{-\infty}^\infty dx^\prime \frac{f(x^\prime)}{z-x^\prime}
\end{equation}
in the complex plane. The function $F(z)$ has the following properties:
\begin{enumerate}

\item it is analytic everywhere in the complex $z$ plane outside the cuts
$\mathcal{S}=\bigcup_{k=1}^M [\alpha_k,\beta_k]$ on the real line

\item  it behaves as $1/z$ when $|z|\to \infty$ since $\int
f(x^\prime)dx^\prime=1$ due to the normalization,

\item  it is real for $z$ real outside the cuts
$\mathcal{S}=\bigcup_{k=1}^M [\alpha_k,\beta_k]$  

\item as one approaches to any point $x$ on the cuts 
$\mathcal{S}=\bigcup_{k=1}^M [\alpha_k,\beta_k]$
on the real axis, $F(x\pm i\epsilon)\stackrel{\epsilon\to 0}{=} g(x) \mp i \pi f(x)$. This
is a consequence of \eqref{prpr}. Thus, $f(x)= -\frac{1}{\pi}{\rm 
Im}[F(x+i\epsilon)]$.

\end{enumerate}

The general theory of analytic functions in the complex plane tells us 
that there is a unique function $F(z)$ which satisfies all the
four properties mentioned above. Thus, if one can make a good guess or 
ansatz for the function 
$F(z)$ and verifies that it satisfies all the above properties, then this
$F(z)$ is unique. Knowing $F(z)$, one can then read off the solution
$f(x)$ using the $4$-th property mentioned above.

In our case, $g(x)=x$, $f(x)=f_c^\star(x)$ and from the simulation
results we already know that there are only two supports for $1/2<c<1$, one 
on the positive side and one on the negative side. To make a good guess
for $F(z)$, let us reexamine the precise form of the solution in the two 
limiting cases $c=1/2$ and $c=1$ 
\begin{align}
\label{f12} f_{1/2}^\star(x) &\propto\sqrt{2-x^2}, 
&\quad \mbox{Wigner's semicircle}\\
\label{f1} f_1^\star(x) &\propto\sqrt{\frac{L_2-x}{x}}[2x+L_2], 
&\quad\mbox{DM \cite{DM}} 
\end{align}
with $L_2=\sqrt{8/3}$. For intermediate values of $c$, we then seek a 
sensible two-support ansatz that 
interpolates 
between \eqref{f12} and \eqref{f1}. A suitable ansatz, that is verified
a posteriori, is 
\begin{equation}\label{ansatz}
f_c^\star(x)=\frac{1}{\pi\sqrt{a}}\sqrt{\frac{L-x}{x}}\sqrt{(ax+L)(x+bL)}
\end{equation}
which has support over $x\in [-L/a,-bL]\cup [0,L]$.
The unknown parameters $a,b,L$ depend on $c$ in such a way that for 
$c\to 
1/2$, $a\to 1,b\to 0, L\to \sqrt{2}$ and for $c\to 1$, $a\to 2,b\to
1/2,L\to\sqrt{8/3}$.
We can then make the following guess for the function $F(z)$, valid everywhere
in the complex plane $z$, except on the cuts $x\in [-L/a,-bL]\cup [0,L]$
on the real axis
\begin{equation}
F(z)=  
z-\sqrt{\frac{z-L}{z}}\sqrt{(z+L/a)(z+bL)}. 
\label{newf}
\end{equation}
It is easy to check that the definition \eqref{newf} indeed satisfies all 
the four properties mentioned above and hence is unique. 
From the 4-th property mentioned above, namely taking the limit
$z\to x+i\epsilon$ with $x\in [-L/a,-bL]\cup [0,L]$, it follows 
that that $f_c^\star(x)$ is indeed given by \eqref{ansatz}.

To fix the
parameters $a$, $b$ and $L$, we will use the 2nd property of $F(z)$
mentioned above, namely that as $|z|\to \infty$, $F(z)\simeq 1/z$.
Expanding $F(z)$ in \eqref{newf} for large $z$ we get
\begin{align}
\nonumber F(z) 
&=z-z\left(1-\frac{L}{z}\right)^{1/2} 
\left(1+\frac{L}{az}\right)^{1/2}\left(1+\frac{bL}{z}\right)^{1/2}\\
&=z-z\left[1+\frac{L}{2}\left(\frac{1}{a}-1+b\right)
\frac{1}{z}+\frac{D(a,b,L)}{z^2} + O(z^{-3})\right]
\end{align}
where
\begin{equation}
D(a,b,L)=\frac{L^2}{8a^2}\left[1+a(2-2b+a(1+b)^2)\right]
\end{equation}
Imposing the exact asymptotic decay $F(z)\simeq 1/z$ for large $|z|$, 
we immediately get the two conditions
\begin{align}
\frac{1}{a}-1+b &=0\\
D(a,b,L) &=1 
\end{align} 
which leads to 
\begin{align}
b &=1-\frac{1}{a}\\
L\equiv L(a) &=\frac{a\sqrt{2}}{\sqrt{a^2-a+1}} \label{eq:L(a)}
\end{align}   
as stated in \eqref{solutionforf}. 
Thus, we are left with only one unknown parameter $a$. This is
fixed from the 
normalization condition 
$\int_{0}^{L(a)}f_c^\star(x)dx= c$ leading to \eqref{normcond} which
determines $a$ implicitly as a function $c$.

\subsection{Method II: proof of \eqref{solutionforf} via double iteration
of the Tricomi solution}

While the method (I) presented in the previous subsection, for finding
the solution with two disconnected supports of the integral equation
\eqref{prpr} with $g(x)=x$, is rather elegant
it has the drawback that one has to make a judicious guess for the
function $F(z)$. It is thus desirable to find a method where one
does not need to guess. We show in this subsection that indeed it
is possible to obtain an explicit two-support solution to \eqref{prpr}
without making an a priori guess. The main idea behind this new method (II)
is to actually use
the Tricomi single support solution twice. Let us first outline below
the basic principle behind this idea which turns out to be rather general
and works for arbitrary source function $g(x)$ in \eqref{prpr}.

We consider again the integral equation
\begin{equation}
\label{saddle1}
g(x)={\rm Pr}\int_{\mathcal{S}} 
dx^\prime \frac{f_c^\star(x^\prime)}{x-x^\prime}
\end{equation}
where $f_c^\star(x)$ is assumed to have nonzero solution over two connected
components $\mathcal{S}=[l_1,l_2]\cup[L_1,L_2]$, with $l_1 \leq l_2\leq 0 
\leq L_1\leq L_2$. Note that the equation \eqref{saddle1} holds
for $x\in [l_1,l_2]$ and also for $x\in [L_1,L_2]$. Let us write
the solution $f_c^\star(x)$ as
\begin{equation}
\label{2dens1}
f_c^\star(x)=\left\{\begin{array}{lll}
f_c^1(x)&  {\rm for}& x\in [l_1,l_2]\\
f_c^2(x)&{\rm for}& x\in [L_1,L_2]
\end{array}\right.
\end{equation}
Then \eqref{saddle1} can be divided into two parts (respectively for the 
left and the right supports) and rewritten as
\begin{align}
g(x) & = \int_{l_1}^{l_2} dx^\prime \frac{f_c^1(x^\prime)}{x-x^\prime}  
+{\rm Pr}\int_{L_1}^{L_2} dx^\prime \frac{f_c^2(x^\prime)}{x-x^\prime}
,\quad\mbox{for } x\in [L_1,L_2] \label{rightsup1} \\
g(x) & = {\rm Pr}\int_{l_1}^{l_2} dx^\prime 
\frac{f_c^1(x^\prime)}{x-x^\prime}  
+\int_{L_1}^{L_2} dx^\prime \frac{f_c^2(x^\prime)}{x-x^\prime}
,\quad\mbox{for } x\in [l_1,l_2] \label{leftsup1} 
\end{align}
Note that for $x\in [L_1,L_2]$, the integral over $[l_1,l_2]$ becomes an
ordinary integral (as there is no pole and we can drop the ${\rm Pr}$) 
and similarly for the other side. 

The main idea then is to eliminate
say $f_c^2(x)$ from these two equations and obtain a single integral
equation for $f_c^1(x)$. This is carried out in the following way.
For $x\in [L_1,L_2]$, \eqref{rightsup1} can be rewritten as
\begin{equation}
{\tilde g}(x)= g(x)-\int_{l_1}^{l_2} dx^\prime 
\frac{f_c^1(x^\prime)}{x-x^\prime}={\rm Pr}\int_{L_1}^{L_2} dx^\prime 
\frac{f_c^2(x^\prime)}{x-x^\prime}.
\label{effect1}
\end{equation}
The solution $f_c^2(x)$ has a single support over $[L_1,L_2]$. Hence
we can now use the explicit Tricomi solution \eqref{formulatricomi}
(replacing $g(x)$ in \eqref{formulatricomi} by the new effective   
source function ${\tilde g}(x)$) to express $f_c^2(x)$ (for $x\in [L_1,L_2]$)
as a functional of $f_c^1(y)$ where $y\in [l_1,l_2]$. Next, we use 
this explicit solution for $f_c^2(x)$ in the second equation \eqref{leftsup1}
and thus obtain a single integral equation involving $f_c^1(x)$. It turns
out that for arbitrary $g(x)$, this integral equation 
for $f_c^1(x)$ can be recast, with a suitable multiplicative factor, in
the same form as \eqref{intequation1} and since $f_c^1(x)$ has only a single 
support
over $[l_1,l_2]$, one can again use the Tricomi solution 
\eqref{formulatricomi} to 
explicitly obtain $f_c^1(x)$. This is the general programme. 
Below we show how the steps actually work out. Even though 
the method is quite general and works for arbitrary $g(x)$, let
us focus below on our specific case $g(x)=x$ just for simplicity.

Our basic saddle point equation reads 
\begin{equation}
\label{saddle}
 x={\rm Pr}\int_{\mathcal{S}} dx^\prime \frac{f_c^\star(x^\prime)}{x-x^\prime}
\end{equation}
where the density $f_c^\star (x)$ must also satisfy the two constraints 
\eqref{n1bis} and \eqref{n2bis}:
\begin{equation}
\label{constraints}
\int_{-\infty}^\infty dx f_{c}^\star(x) =1
\;\; {\rm and} \;\; \int_{0}^\infty dx f_{c}^\star(x) =c.
\end{equation}
The solution $f_c^\star (x)$ is expected to have support over two
disconnected components  
$\mathcal{S}=[l_1,l_2]\cup[L_1,L_2]$, with $l_1 \leq l_2\leq 0 \leq L_1\leq L_2$.
For consistency, we expect $f_c^\star(l_1)=0=f_c^\star(L_2)$.
We also expect
$f_c^\star(L_1)=0$ if $L_1>0$ (or otherwise $L_1=0$
with no constraint on $f_c^\star(L_1)$), and similarly $f_c^\star(l_2)=0$ 
if $l_2<0$.
We divide $f_c^\star(x)$ into two parts as in \eqref{2dens1}.
The constraints thus become:
\begin{equation}
\label{2constraints}
 \int_{l_1}^{l_2} dx f_{c}^1(x)+\int_{L_1}^{L_2} dx f_c^2(x) =1
\;\; {\rm and}\;\; \int_{L_1}^{L_2} f_{c}^2(x) =c
\end{equation}

We then apply Tricomi's theorem \eqref{formulatricomi} to 
\eqref{effect1} with $g(x)=x$ to
determine
$f_c^2(y)$ on the interval $y\in [L_1,L_2]$ and obtain 
\begin{align}
\nonumber f_c^2(y)&=\frac{1}{\pi^2 \sqrt{y-L_1}\sqrt{L_2-y}}
\left[ \pi c+{\rm Pr}\int_{L_1}^{L_2}du \frac{\sqrt{u-L_1}\sqrt{L_2-u}}{u-y} 
\left( u+
\int_{l_1}^{l_2} dt \frac{f_c^1(t)}{t-u}\right)
\right]=\\
&=\frac{1}{\pi \sqrt{y-L_1}\sqrt{L_2-y}}
\left[1+\frac{(L_2-L_1)^2+4 (L_1+L_2)y-8 y^2}{8}
+\int_{l_1}^{l_2}dt f_c^1(t) \frac{\sqrt{L_1-t}\sqrt{L_2-t}}{t-y}
 \right] \label{rightsup2}
\end{align}
where we have used the following result:
\begin{equation}
{\rm Pr}\int_{L_1}^{L_2}du \frac{\sqrt{u-L_1}\sqrt{L_2-u}}{(u-y)(t-u)}=
\pi \left(1+ \frac{\sqrt{L_1-t}\sqrt{L_2-t}}{t-y}\right)
\end{equation}
and 
\begin{equation}
\int dx f_c^1(x)=1-\int dx f_c^2(x)=1-c
\end{equation}

As explained above, we expect $f_c^\star(L_2)=0$.
Thus
\begin{equation}
\label{constr3}
1+\frac{L_1^2+2 L_1 L_2-3 L_2^2}{8}
+\int_{l_1}^{l_2}dt f_c^1(t) \frac{\sqrt{L_1-t}\sqrt{L_2-t}}{t-L_2}=0.
\end{equation}
Multiplying $f_c^2(y)$ by $\pi \sqrt{(y-L_1)(L_2-y)}$ in \eqref{rightsup2}
and subtracting \eqref{constr3} from it gives a rather compact expression
\begin{equation}
\label{fc2}\hspace{-0.2cm}
f_c^2(y)=\frac{1}{\pi}\sqrt{\frac{L_2-y}{y-L_1}}\left[
\frac{L_2-L_1}{2}+y+\int_{l_1}^{l_2} dt \frac{f_c^1(t)}{t-y}
\sqrt{\frac{L_1-t}{L_2-t}}
\right]\;\;{\rm for} \; y\in [L_1,L_2]
\end{equation}

Next we substitute this expression of $f_c^2(x)$ 
in the saddle point equation \eqref{leftsup1} valid over
the left support $[l_1,l_2]$ (with $g(x)=x$).
The resulting integrals can be carried out explicitly.
We need to use the following integral
\begin{equation}
\frac{1}{\pi}\int_{L_1}^{L_2} \frac{dy}{x-y} \sqrt{\frac{L_2-y}{y-L_1}}=
1-\sqrt{\frac{L_2-x}{L_1-x}}
\label{int1}
\end{equation}
valid for $x<L_1<L_2$. 
After a few steps of algebra we get
\begin{align}
& x-{\rm Pr}\int_{l_1}^{l_2}dt \frac{f_c^1(t)}{x-t}=\int_{L_1}^{L_2}dy 
\frac{f_c^2(y)}{x-y} =\nonumber\\
&= x-\frac{L_1+L_2}{2}\sqrt{\frac{L_2-x}{L_1-x}}+\sqrt{L_1-x}\sqrt{L_2-x}
-{\rm Pr}\int_{l_1}^{l_2}dt \frac{f_c^1(t)}{x-t} 
-{\rm Pr}\int_{l_1}^{l_2}dt \frac{f_c^1(t)}{t-x}
\sqrt{\frac{L_1-t}{L_2-t}}\sqrt{\frac{L_2-x}{L_1-x}}.
\end{align}
Cancellations of terms from both sides then lead us to the following integral
equation for $f_c^1(x)$ for $x\in [l_1,l_2]$
\begin{equation}
\label{newSaddle}
{\rm Pr}\int_{l_1}^{l_2}dt \frac{f_c^1(t)}{t-x}\sqrt{\frac{L_1-t}{L_2-t}}=
\frac{L_1-L_2}{2}-x.
\end{equation}
Defining $\tilde{f_c^1}(x)\equiv f_c^1(x)\sqrt{\frac{L_1-x}{L_2-x}}$,
we get an integral equation over $[l_1,l_2]$ 
\begin{equation}
{\rm Pr}\int_{l_1}^{l_2}dt   
\frac{\tilde{f_c^1}(t)}{t-x}=\frac{L_1-L_2}{2}-x
\label{newSaddle1}
\end{equation}
which, fortunately, has the same form as the original single-support
saddle point equation \eqref{intequation1} with the source function 
$g(x)=(L_1-L_2)/2-x$. This can be inverted explicitly using
\eqref{formulatricomi}. 
Enforcing the constraint $f_c^1(l_1)=0$, we get
\begin{equation}\label{constr4}
\int_{l_1}^{\l_2} \tilde{f_c^1}(x)dx + \frac{(L_2-L_1)(l_2-l_1)}{4}+
\frac{l_2^2-3l_1^2+2 l_1 l_2}{8}=0
\end{equation}
Using this in the Tricomi formula \eqref{formulatricomi} finally
gives us a rather explicit solution   
\begin{equation}
\label{fc1}
f_c^1(x)=\frac{1}{\pi}\sqrt{\frac{x-l_1}{l_2-x}}\sqrt{\frac{L_2-x}{L_1-x}}
\left[\frac{L_1-L_2}{2}+\frac{l_2-l_1}{2}-x \right]
\;\;
{\rm for} \;\;
l_1\leq x<l_2
\end{equation}
We can now replace $f_c^1(x)$ in the expression of $f_c^2(x)$
given in Eq. \eqref{fc2}.
Finally we get the expression of the density $f_c^\star(x)$ ($f_c^\star(x)=f_c^1(x)$
on $[l_1,l_2]$ and $f_c^\star(x)=f_c^2(x)$ on $[L_1,L_2]$):
\begin{equation}
\label{fc}
f_c^\star(x)=\frac{1}{\pi}\sqrt{\left(\frac{x-l_1}{l_2-x}\right)\,
\left( \frac{L_2-x}{L_1-x}\right)}
\:\left|\frac{L_1-L_2}{2}+\frac{l_2-l_1}{2}-x \right|
\;\;
{\rm for} \;\;
x\in[l_1,l_2]\cup [L_1,L_2]
\end{equation}

So far we have used two physical conditions $f_c^1(l_1)=0$ and $f_c^2(L_2)=0$
which are evidently manifest in the explicit solution \eqref{fc}. 
Substituting in \eqref{constr3}
the expression
of $f_c^\star (x)$ from Eq. \eqref{fc}, we get
an identity for the edge points of the support
\begin{equation}
\label{constrB}
1+\frac{L_1^2+2 L_1 L_2 -3 L_2^2}{8}+\left(\frac{l_2-l_1}{8}\right)\left(
3 l_1+l_2+2 L_2-2 L_1 \right)=0
\end{equation}
In addition, we have one more condition $\int_{L_1}^{L_2} f_c^2(x)dx=c$.
Thus we have four unknowns $l_1$, $l_2$, $L_1$ and $L_2$ and two
conditions mentioned above. To determine all the constants, we need
to impose some additional conditions at the other two edges $x=l_2$
and $x=L_1$. With these conditions imposed, one obtains a unique solution
for a given value of $c$ as demonstrated below. 

It is clear 
we must have either $L_1=0$, or $L_1>0$ (but with $f_c^\star(L_1)=0$).
Similarly, we must also have either $l_2=0$, or $l_2<0$ 
(with $f_c^\star(l_2)=0$).

\begin{itemize}

\item {\bf First case:} $l_2=0=L_1$.
\\
Eq. \eqref{constrB} gives
$ 8=3 L_2^2+3 l_1^2+2 l_1 L_2$. Thus:
\begin{equation*}
f_c^\star(x)=\frac{1}{\pi\, }\sqrt{(x-l_1) (L_2-x)}
\:\frac{\left|\frac{l_1+L_2}{2}+x \right|}{|x|}
\;\;
{\rm for} \;\;
x\in[l_1,0[\cup]0,L_2]
\end{equation*}
The last constraint $\int_{0}^{L_2} f_c^\star(x)dx =c$
implies that $f_c^\star(x)$ is integrable in zero, thus
$l_1+L_2=0$.
Finally, using Eq. \eqref{constrB} we get $L_2=-l_1=\sqrt{2}$ and:
 \begin{equation}\label{semicircle}
f_c^\star(x)=\frac{1}{\pi\, }\sqrt{2-x^2}
\;\;
{\rm for} \;\;
x\in[-\sqrt{2},\sqrt{2}]
\end{equation}
and we recover the Wigner semicircle law,
having a single support $[-\sqrt{2},\sqrt{2}]$.
Note that in this case $c=\int_{0}^{\sqrt{2}}dx  f_c^\star(x) =\frac{1}{2}$
already is fixed. Thus, this solution is valid only for $c=1/2$.
\item {\bf Second case:} $l_2<0$ with $f_c^\star(l_2)=0$ and $L_1=0$.
\\
In this case, the density has a  support over
$[l_1,l_2]\cup]0,L_2]$.
We get:
\begin{equation}
\label{fcVivo}
f_c^\star(x)=\frac{1}{\pi}\sqrt{\frac{(x-l_1)(x-l_2)(L_2-x)}{x}\,}
\;\;
{\rm for} \;\;
x\in [l_1,l_2]\cup]0,L_2]
\end{equation}
with $L_2=-(l_1+l_2)$ (because $f_c^\star(l_2)=0$)
and $1+\frac{-3 L_2^2}{8}+\left(\frac{l_2-l_1}{8}\right)\left(
3 l_1+l_2+2 L_2 \right)=0$ (Eq. \eqref{constrB}).
Let us define $a=-L_2/l_1$.
We readily obtain the claimed solution \eqref{solutionforf}:
\begin{equation}
\label{vivoBounds}
L_2=L(a)=\frac{a \sqrt{2}}{\sqrt{a^2-a+1}}\;\;
{\rm and} \;\; l_1=-\frac{L_2}{a}\;\; {\rm and} \;\;
l_2=-L_2 \left(1-\frac{1}{a} \right)
\end{equation}
As $l_2 \leq 0$ and $l_1\leq l_2$, we have: $1\leq a\leq 2$.
Because of the last constraint $\int_{0}^{L_2}f_c^\star(x)dx=c$,
 the parameter $a$ must also
satisfy the following  equation:
\begin{equation}
\label{eqA}
\int_0^1 dy \sqrt{\frac{1-y}{y}}\, \sqrt{y^2+y+\frac{a-1}{a^2}}
=\frac{\pi}{2}\left( 1-\frac{a-1}{a^2}\right)\:  c
\end{equation}
in complete agreement with \eqref{normcond}.

\item {\bf Third case:} $L_1>0$ with $f_c^\star(L_1)=0$ and $l_2=0$.
\\
This is the exact symmetric of the second case.
It corresponds to $c<1/2$.
\item {\bf Fourth case:} $l_2<0$ with $f_c^\star(l_2)=0$ and $L_1>0$
with $f_c^\star(L_1)=0$.
\\
The constraints   $f_c^\star(l_2)=0$ and $f_c^\star(L_1)=0$
give respectively $L_2-L_1=-(l_1+l_2)$
and $L_1+L_2=l_2-l_1$. Thus $L_2=-l_1$ and $L_1=l_2$.
As $l_2< 0 < L_1$, this 
case is impossible.

\end{itemize}

In conclusion, there is only one unique solution (case 2 above) which
is valid for all $1/2\le c\le 1$ and in the limiting case $c=1/2$ this
solution coincides with the first case above that is valid only for $c=1/2$.

\subsection{Evaluation of the action and derivation of $\Phi(c)$}

Having computed explicitly the saddle point solution $f_c^\star(x)$
in Eqs. \eqref{solutionforf}-\eqref{normcond}, the next step
is to evaluate the saddle point action $S[f_c^\star(x)]$ where
the action $S[f_c(x)]$ is given in \eqref{action}. This will then
provide the expression for the large deviation function $\Phi(c)$
associated with the index distribution in \eqref{largedevlaw}
\begin{equation}
\Phi(c) = \frac{1}{2} S[f_c^\star(x)]- \frac{(3+2\log(2))}{8}.
\label{phic11}
\end{equation}
 
Upon substituting the saddle point solution $f_c^\star(x)$ in the 
action \eqref{action}, one gets:
\begin{equation}
S[f_c^\star(x)]= \int_{-\infty}^{\infty} x^2 f_c^\star(x) dx
-\int_{-\infty}^{\infty}\int_{-\infty}^{\infty} f_c^\star(x) f_c^\star(x') 
\log|x-x'| 
dx\, dx'.
\label{action1}
\end{equation}
By construction, the saddle point solution $f_c^\star(x)$ automatically
satisfies the two constraints and hence the terms involving the two Lagrange 
multipliers drop out in \eqref{action}. One can directly substitute
the explicit expression of $f_c^\star(x)$ from \eqref{solutionforf}
to evaluate the double integral in \eqref{action1}. However, this
is a bit cumbersome. It turns out
to be convenient to use a slightly different trick. Note that
$f_c^\star(x)$ satisfies the saddle point equation 
\begin{equation}
x^2 + A_1 \theta(x) + A_2 = 
2\int_{-\infty}^{\infty} f_c^\star(x') \log|x-x'|\, dx'
\label{inteq1}  
\end{equation}
The important point is that this equation is valid for all $x$ where
the solution $f(x)$ is nonzero, i.e., for all $x\in [l_1,l_2]\cup [0,L(a)]$
where $l_1=-L(a)/a$, $l_2=-(1-1/a)L(a)$ and $L(a)$ is given in 
\eqref{La}. 

To evaluate the action, we multiply \eqref{inteq1} by $f_c^\star(x)$ and 
integrate over all $x$. Using the two normalization conditions: (i) 
$\int_{-\infty}^{\infty} f_c^\star(x) dx=1$ and (ii) $\int_0^{\infty} 
f_c^\star(x)dx=c$ we get 
\begin{equation} 
\int_{-\infty}^{\infty}\int_{-\infty}^{\infty} f_c^\star(x) f_c^\star(x') 
\log|x-x'| dx\, dx'= 
\frac{1}{2}\left[\int_{-\infty}^{\infty} x^2 f_c^\star(x) dx 
+ A_1 c + A_2\right]. 
\label{doubleint} 
\end{equation} 
Substituting this result 
in the action \eqref{action1} gives 
\begin{equation} 
S[f_c^\star(x)]= 
\frac{1}{2}\left[\int_{-\infty}^{\infty} x^2 f_c^\star(x) dx - A_1 c -A_2\right] 
\label{action2} 
\end{equation} Denoting $\mu_2= \int_{-\infty}^{\infty} x^2 
f_c^\star(x) dx$ we get from \eqref{phic11}  
\begin{equation} 
\Phi(c) = 
-\frac{1}{4}\left[(3/2-\mu_2)+\log(2) + A_1 c+ A_2\right] 
\label{phic2} 
\end{equation} 
It then remains to evaluate $\mu_2$ and the Lagrange multipliers 
$A_1$ and $A_2$.

To determine the Lagrange multipliers we proceed as follows.
Let us recall the function $F(z)$ defined in \eqref{Frho} for all
$z$ in the complex plane except on the real cuts $x\in [-L(a)/a,-(1-1/a)L(a)]\cup 
[0,L(a)]$. Setting $z=x$ real, but outside these two cuts, and $L\equiv L(a)$ we can make a
large $x$ expansion
\begin{equation}
F(x) = \int \frac{f_c^\star(x')}{x-x'}\, dx' = \sum_{n=0}^{\infty} 
\frac{\mu_n}{x^{n+1}}
\label{Fx1}
\end{equation}
where $\mu_n = \int f_c^\star(x) x^n dx$ is the $n$-th moment. From the
explicit solution of $f_c^\star(x)$ in Eq. (\ref{solutionforf}) one can check
that $\mu_0=1$ and also $\mu_2=1/2$ (independent of $c$).

In addition, for real $x>L$, we have from Eq. \eqref{newf}
\begin{equation}
F(x) = x- \sqrt{\frac{(x-L)}{x}\, 
\left(x+\frac{L}{a}\right)\left(x+\left(1-\frac{1}{a}\right)L\right)}  
\label{Fxr}
\end{equation}
On the other hand for $x<-L/a$ (on the real line to the left of the edge
$-L/a$ of the left support), the function $F(x)$ has the form
\begin{equation}
F(x)= x + \sqrt{\frac{(x-L)}{x}\, 
\left(x+\frac{L}{a}\right)\left(x+\left(1-\frac{1}{a}\right)L\right)}
\label{Fxl}
\end{equation}
where the square-root is always chosen to be positive. Note that
with this choice in \eqref{Fxl}, $F(x)\approx 1/x$ for large negative $x$.

To determine $A_1$ and $A_2$, we need to choose a value of $x$ in 
\eqref{inteq1} such that it belongs to either of the two supports.
Choosing $x=L$ and $x=-L/a$ gives the following two equations
\begin{align}
L^2 + A_1 + A_2 &= 2 \int f_c^\star(x') \log(L-x')\, dx' \label{lm1} \\
L^2/a^2 + A_2 & = 2 \int f_c^\star(x') \log(x'+L/a)\, dx' \label{lm2}
\end{align}
where the integral runs only over the supports. Writing
$\log(L-x')=\log(L)+\log(1-x'/L)$, expanding the logarithm in a series
and using the definition of $\mu_n$ we get from \eqref{lm1} 
\begin{equation}
L^2 + A_1 +A_2 = 2 \log(L) - 2 \sum_{n=1}^{\infty} \frac{\mu_n}{n L^n}
\label{con1}
\end{equation}  
Similarly, in Eq. (\ref{lm2}) we write $\log(x'+L/a)= \log(L/a) +
\log(1+ax'/L)$ and expand the logarithm in a series to get  
\begin{equation}
L^2/a^2+A_2 = 2 \log(L/a) - 2\sum_{n=1}^{\infty} \frac{(-a)^n \mu_n}{nL^n}
\label{con2}
\end{equation}

We can then determine $A_1$ and $A_2$ in terms of $\mu_n$ by solving the
two linear equations (\ref{con1}) and (\ref{con2}).
It is actually
convenient to express the series involving $\mu_n$ in terms of the following
integrals. Using Eq. (\ref{Fx1}) and using $\mu_0=1$ we get,
\begin{equation}
F(x)-\frac{1}{x} = \sum_{n=1}^{\infty} \frac{\mu_n}{x^{n+1}}
\label{ser1}
\end{equation}
Let us first consider the regime $x\ge L$. Here, let us define
\begin{equation}
W_1(x) = F(x) -\frac{1}{x} = x-\frac{1}{x}- \sqrt{\frac{(x-L)}{x}\,
\left(x+\frac{L}{a}\right)\left(x+\left(1-\frac{1}{a}\right)L\right)}
\label{w1x}
\end{equation}
where we have used the definition of $F(x)$ in Eq. (\ref{Fxr}).
Integrating Eq. (\ref{ser1}) over  $[L,\infty]$ gives
\begin{equation}
\int_L^{\infty} W_1(x) dx = \sum_{n=1}^{\infty}\frac{ \mu_n}{nL^n}
\label{sum1}
\end{equation} 

\begin{figure}[htbp]
\includegraphics[height=6.0cm]{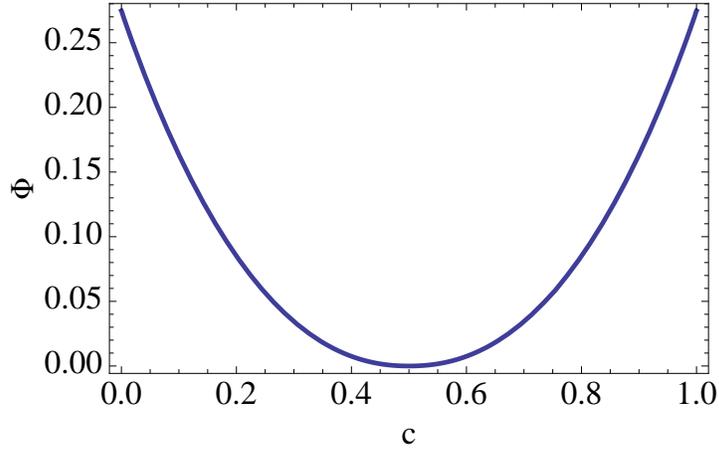}\hspace{0.5cm}
\caption{The large deviation function $\Phi(c)$ in \eqref{phic1}.}
\label{fig:phic}
\end{figure}

Next we consider the regime $x\le -L/a$. Here we use the definition
of $F(x)$ in Eq. (\ref{Fxl}). Integrating Eq. (\ref{ser1}) over $[-\infty,
-L/a]$ gives  
\begin{equation}
\int_{-\infty}^{-L/a} \left[F(x)-\frac{1}{x}\right]dx = - 
\sum_{n=1}^{\infty}\frac{(-a)^n 
\mu_n}{n
L^n}
\label{sum2}
\end{equation}
It is convenient to make a change of variable $x\to -x$ on the l.h.s
of Eq. (\ref{sum2}). Using the definition of $F(x)$ in Eq. (\ref{Fxl})
this finally gives
\begin{equation}
\int_{L/a}^{\infty} W_2(x) dx = \sum_{n=1}^{\infty}\frac{(-a)^n \mu_n}{n
L^n}
\label{sum3}
\end{equation}
where $W_2(x)$ is given by
\begin{equation}
W_2(x) = x-\frac{1}{x} -
\sqrt{\frac{(x+L)}{x}\left(x-\frac{L}{a}\right)
\left(x-\left(1-\frac{1}{a}\right)L\right)}
\label{w2}
\end{equation}
Next we insert the expressions of the two sums from Eqs. (\ref{sum1}) and
(\ref{sum3})
in the two linear equations (\ref{con1}) and (\ref{con2}), solve for $A_1$
and $A_2$ and then substitute them in Eq. (\ref{phic2}). This then yields
the main result 
\begin{equation}
\Phi(c) = \frac{1}{4}[L^2-1-\log(2L^2)] + \frac{(1-c)}{2}\,\log(a)
-\frac{(1-c)(a^2-1)}{4a^2}\,L^2 + \frac{c}{2}\int_{L}^{\infty}W_1(x) 
dx
+\frac{(1-c)}{2}\int_{L/a}^{\infty} W_2(x)dx
\label{phic1}   
\end{equation}
where $W_1(x)$ and $W_2(x)$ are defined respectively in Eqs. \eqref{w1x}
and \eqref{w2}. Unfortunately the two integrals are difficult
to compute analytically. However, they can be easily evaluated by
Mathematica.  
A plot of this function is provided in Fig. (\ref{fig:phic}). This final form 
turns out 
to be the most convenient
one for carrying out the asymptotic expansion near $c=1/2$ in the
next subsection.

\subsection{Asymptotic Expansion of $\Phi(c)$ near $c=1/2$}

We now expand $\Phi(c)$ in Eq. (\ref{phic1}) for $c$ close to $1/2$.
We set $c=1/2+\delta$ with $\delta\ge 0$ being small. Let us also
define the parameter $\epsilon$ by
\begin{equation}
\frac{(a-1)}{a^2}= \epsilon
\label{eps1}
\end{equation}
When $c\to 1/2$, $a\to 1$ from Eq. (\ref{normcond}), hence $\epsilon$ is a
small parameter for $c$ close to $1/2$. It follows from Eq. (\ref{eps1})
that
\begin{equation}
a= \frac{1-\sqrt{1-4\epsilon}}{2\epsilon}
\label{adef1}
\end{equation}
where we have chosen the root that gives $a\to 1$ as $\epsilon\to 0$.
It also follows from Eq. (\ref{La}) that
\begin{equation}
L^2 = \frac{2}{1-\epsilon}
\label{Ldef1}
\end{equation}

Let us first establish a relation between $\delta$ and $\epsilon$ when
both are small. Eq. (\ref{normcond}), in terms of $\epsilon$ and $\delta$,
can be recast as
\begin{equation}
J(\epsilon)= \int_0^1 dy\, \sqrt{\frac{1-y}{y}}\,\sqrt{y^2+y+
\epsilon}= \frac{\pi}{2} (1-\epsilon)(1/2+\delta).
\label{delta1}
\end{equation}
Let us first analyze the integral on the l.h.s of Eq. (\ref{delta1}).
To find its asymptotic behavior for small $\epsilon$, we first  note
that $J(0)= \pi/4$. Next, taking a derivative with respect to
$\epsilon$ gives
\begin{equation}
J'(\epsilon)= \frac{1}{2}\int_0^1 dy\, \sqrt{\frac{1-y}{y}}\, 
\frac{1}{\sqrt{y^2+y+\epsilon}}
\label{deps1}
\end{equation}  
Make a change of variable $y=\epsilon z$ in the integral and take the limit
$\epsilon\to 0$. To leading order in small $\epsilon$ one easily finds
\begin{equation}
J'(\epsilon)= -\frac{1}{2} \log (\epsilon)
\label{deps2}
\end{equation}
Integrating and using $J(0)=\pi/4$ one then finds for small $\epsilon$
\begin{equation}
J(\epsilon) = \frac{\pi}{4} - \frac{1}{2} \epsilon \log(\epsilon) +\ldots
\label{Jasymp1}
\end{equation}
Comparing the left and the right hand side of Eq. (\ref{delta1}) then
gives, to leading order in small $\epsilon$
\begin{equation}
\delta= -\frac{1}{\pi} \epsilon \log (\epsilon)
\label{deltaasymp}
\end{equation}
Inverting Eq. (\ref{deltaasymp}), one can express $\epsilon$ as a function of 
$\delta$ and to leading order for small $\delta$ one gets
\begin{equation}
\epsilon= \frac{\pi \delta}{-\log(\delta)}
\label{epsasymp}
\end{equation}

We are ready to expand $\Phi(c)$ in Eq. (\ref{phic1}) for small
$\delta$ (or equivalently for small $\epsilon$). There are $5$ terms
on the right hand side of Eq. (\ref{phic1}). We expand each of them
separately.

The first term gives, upon using Eq. (\ref{Ldef1})
\begin{equation}
T_1= \frac{1}{4}\left[L^2-1-\log(2L^2)\right]= \frac{1}{4}[1-\log(4)] 
+\frac{1}{4}{\epsilon} + \frac{3}{8}\epsilon^2 + \mathcal{O}(\epsilon^3).
\label{t1}
\end{equation}

The second term, upon using $c=1/2+\delta$ and $a$ from Eq. (\ref{adef1})
and expanding for small $\epsilon$ 
\begin{equation}
T_2 = \frac{1-c}{2}\log(a) = \frac{1}{4}\epsilon -\frac{1}{2} \epsilon\delta 
+\mathcal{O}(\epsilon^2)
\label{t2}
\end{equation}

The third term gives
\begin{equation}
T_3= -\frac{(1-c)}{4}(1-1/a^2)L^2 = -\frac{1}{2}\epsilon+ \epsilon\delta + 
\mathcal{O}(\epsilon^2)
\label{t3}
\end{equation}

The fourth term gives:
\begin{align}
T_4 &=  \frac{c}{2}\int_{L}^{\infty}dx\, \left[x-\frac{1}{x} -
\sqrt{\frac{(x-L)}{x}\left(x+\frac{L}{a}\right)
\left(x+\left(1-\frac{1}{a}\right)L\right)}\right] \nonumber \\
&= \frac{1}{8} [-1+\log(4)]+\frac{\pi-1}{8} \epsilon 
+\frac{(-1+\log(4))}{4}\delta + \frac{\pi-1}{4}\epsilon\delta + \mathcal{O}(\epsilon^2)
\label{t4}
\end{align}

Similarly, the fifth term gives:
\begin{align}
T_5 &= \frac{(1-c)}{2}\int_{L/a}^{\infty} dx\, \left[x-\frac{1}{x} -
\sqrt{\frac{(x+L)}{x}\left(x-\frac{L}{a}\right)
\left(x-\left(1-\frac{1}{a}\right)L\right)}\right] \nonumber \\
&= \frac{1}{8}[-1+\log(4)]-\frac{\pi-1}{8} 
\epsilon-\frac{(-1+\log(4))}{4}\delta+ \frac{\pi-1}{4}\epsilon\delta + 
\mathcal{O}(\epsilon^2)
\label{t5}
\end{align}

Adding the five terms one gets, to leading order,
\begin{equation}
\Phi(c=1/2+\delta)= T_1+T_2+T_3+T_4+T_5= \frac{\pi}{2} \epsilon \delta 
+\mathcal{O}(\epsilon^2)
\label{phicas1}
\end{equation}
Using the expression of $\epsilon$ as a function of $\delta$ from Eq. 
(\ref{epsasymp}) then gives our leading order result for small $\delta$
\begin{equation}
\Phi(c=1/2+\delta)\simeq -\frac{\pi^2}{2} \frac{\delta^2}{\log \delta}
\label{phicas2}
\end{equation}
Substituting this result in Eq. (\ref{largedevlaw}) we then get, for 
$c=1/2+\delta$
with $\delta$ small (note that by symmetry one can similarly obtain
the form of the function for $\delta<0$ also)
\begin{equation}
{\mathcal P}((1/2+\delta)N,N)\approx \exp\left[-{\beta}{\pi^2}N^2\, 
\frac{\delta^2}{-2\log(|\delta|)}\right].
\label{large2}
\end{equation}
Resetting $\delta=(\mathcal{N}_+-N/2)/N$ and assuming $(\mathcal{N}_+-N/2)<< 
N$ one gets the Gaussian distribution in the large $N$ limit
\begin{equation}
{\mathcal P}(\mathcal{N}_+, N)\approx \exp\left[-
\frac{\beta \pi^2}{2\log
(N)}\left(\mathcal{N}_+-N/2\right)^2\right]
\label{gauss11}
\end{equation}
from which one can read off the variance  
for large $N$ and for all $\beta$
\begin{equation}
\Delta(N)= \Big\langle \left(\mathcal{N}_{+}-\frac{N}{2}\right)^2\Big\rangle 
\simeq \frac{1}{\beta 
\pi^2} \log(N)+\mathcal{O}(1)
\label{variance}
\end{equation}
This result is in agreement with that of 
Cavagna \textit{et al.} \cite{CGG} for $\beta=1$.

\section{Numerical Simulations}
\label{sec:simulnum}

In this section, we explain how to compute numerically the 
index distribution for a Gaussian random matrix ensemble and to compare the results with
analytical predictions.
The joint distribution of the $N$ eigenvalues
of a $N\times N$ Gaussian random matrix with Dyson index $\beta$
is given in Eq. \eqref{eq:jpdfEV} by:
\begin{equation}\label{jpdfEV}
\mathcal{P}(\lambda_1,\ldots,\lambda_N)=\frac{1}{Z_N}e^{-\frac{\beta}{2}\sum_{i=1}^N\lambda_i^2}\prod_{j<k}|\lambda_j-\lambda_k|^\beta 
=\frac{1}{Z_N}\: e^{-\beta E\left[\{ \lambda_i\} \right]}
\end{equation}
with $ E\left[\{ \lambda_i\} \right]=\frac{1}{2} \sum_i \lambda_i^2
-\sum_{i<j} \log\left|\lambda_i-\lambda_j \right|$.
The idea is to sample the distribution in Eq. \eqref{jpdfEV}
 using a Metropolis Monte Carlo algorithm
and to construct a histogram of the number of positive 
eigenvalues $\mathcal{N}_{+}=\sum_{i=1}^N \theta(\lambda_i)$.
For large $N$, we expect the distribution
of $\mathcal{N}_{+}$ to be of the form (see Eq. \eqref{ldv0}):
\begin{equation}
\mathcal{P}({\mathcal N}_+=cN,N)\sim\exp\left[-\beta N^2 \Phi(c)\right]
\end{equation}
Therefore we want to construct a histogram
of the rate function $\Phi_{\rm num}(c)\equiv -\frac{\log \mathcal{P}({\mathcal N}_+=cN,N)}{\beta N^2}$ 
and compare with its analytical expression $\Phi(c)$ for large $N$ given in Eq. \eqref{phic1}.

As $\mathcal{N}_{+}$ is a discrete function of the $N$ eigenvalues, it takes integer values between $0$ and $N$.
Numerically it is easier to consider  continuous functions and to come back to $\mathcal{N}_{+}$ only at the end.
Therefore we introduce a smoothed version of the  Heaviside theta function $\theta(\lambda)$ and thus of $\mathcal{N}_{+}$.
Let us define for $\eta >0$:
\begin{equation}
 \theta_{\eta}(\lambda)= \frac{1}{1+e^{-\frac{\lambda}{\eta}}}\;\;\;\; {\rm and} \;\; \;\;
\mathcal{N}_{\eta}=\sum_{i=1}^N \theta_{\eta}(\lambda_i)
\end{equation}
The function $\theta_{\eta}$ increases from $0$ (in the limit $\lambda \rightarrow - \infty$) to $1$
(in the limit $\lambda\rightarrow \infty$). It has the same symmetry with respect to the origin as the Heaviside theta function:
$\theta_{\eta}(-\lambda)=1-\theta_{\eta}(\lambda)$. Thus we have
$\mathcal{P}({\mathcal N}_{\eta}=cN,N)=\mathcal{P}({\mathcal N}_{\eta}=(1-c)N,N)$.
 The parameter $\eta$ gives the width of the jump from $0$ to $1$ and
$\lim_{\eta\rightarrow 0} \theta_{\eta}(\lambda)=\theta(\lambda)$, thus $\mathcal{N}_0=\mathcal{N}_+$.

\subsection{Distribution of $\mathcal{N}_{\eta}$ : non-standard Metropolis algorithm}

In this section, we explain the Metropolis algorithm
and a modified version that allows us to reconstruct numerically the full
distribution of $\mathcal{N}_{\eta}$ for a fixed and large enough value of $\eta$.

\subsubsection{Standard Metropolis algorithm}

We start with an initial configuration of the $\lambda_i's$
(real numbers of order $\sqrt{N}$).
At each step, a small move $\{\lambda_i\} \longrightarrow \{\lambda_i'\}$
is proposed in the configuration space.
In our algorithm, it consists of
 picking at random an eigenvalue $\lambda_j$
and proposing to modify it as  $\lambda_j
\longrightarrow \lambda_j +\epsilon$,
where
$\epsilon$ is a real number drawn from a Gaussian distribution with mean zero
and with a variance that is set to achieve the standard average rejection rate $1/2$. 

The move is accepted with probability
\begin{equation}
\label{DetailedBalance}
p=\min\left(\frac{
\mathcal{P}(\lambda_1',\ldots,\lambda_N')}{
\mathcal{P}(\lambda_1,\ldots,\lambda_N)},1 \right)
=\min\left( e^{-\beta \left(E\left[ \{\lambda_i' \}\right]-
E\left[\{\lambda_i \}\right]\right)},1\right)
 \end{equation}
and rejected with probability $1-p$.
This dynamics enforces the detailed balance 
and ensures that at long times the algorithm reaches  thermal equilibrium
(at inverse ``temperature'' $\beta$) with the correct Boltzmann weight
$e^{-\beta E\left[\{\lambda_i\}\right]}$.

At long times,
 the Metropolis algorithm thus generates
samples of $\{\lambda_i\}$ drawn from the joint distribution
in Eq. \eqref{jpdfEV}. We can start to keep the value of $\mathcal{N}_{\eta}=\sum_{i=1}^N \theta_{\eta}(\lambda_i)$
 for the configurations of eigenvalues generated by the algorithm
(say every ten steps)
and construct a histogram for $\mathcal{N}_{\eta}$.

However, the distribution of $\mathcal{N}_{\eta}$
is expected to be of the form $\mathcal{P}({\mathcal N}_{\eta}=c_{\eta} N,N)\sim\exp\left[-\beta N^2 \Phi_{\eta}(c_{\eta})\right]$
for large $N$
exactly as for $\mathcal{N}_+$, and thus to be highly peaked around its average.
The events in the tails of the distribution are extremely rare. Therefore we can not, with
a standard Metropolis algorithm, explore  in a "reasonable" time
a wide range of values of $\mathcal{N}_{\eta}$. 
We propose below a modified version of the algorithm
that allows us to explore the far left and right tails of the distribution.

\subsubsection{Modified algorithm: conditional probabilities}

We want to explore regions that are far from the mean value of $\mathcal{N}_{\eta}$, ie far from 
$\langle \mathcal{N}_{\eta}\rangle = N/2$ (by symmetry), for
example the far right tail $\mathcal{N}_{\eta}=N c_{\eta}$ with $c_{\eta} > 1/2$.

The idea is thus to force the algorithm to explore the region $c_{\eta}\geq c^{*}$ for
different values of $c^*$. We thus add in the algorithm the constraint $c_{\eta}\geq c^{*}$.
More precisely, we start with an initial configuration that satisfies
$\mathcal{N}_{\eta}=c_{\eta} N\geq N c^{*}$. At each step, the move is rejected
if  $\mathcal{N}_{\eta} < N c^*$.
If  $\mathcal{N}_{\eta} \geq N c^*$, then the move is accepted or rejected 
exactly with the same condition as before (see Eq. \eqref{DetailedBalance}).
Because of the new constraint $c_{\eta}\geq c^{*}$, the moves are rejected  more often
than before. Therefore the variance of the Gaussian distribution $P(\epsilon)$
has to be taken smaller  to achieve the standard rejection rate $1/2$.

We run  the program for several values of $c^*$
and we  construct a histogram of $\mathcal{N}_{\eta}$ for each value $c^*$.
This gives the conditional probability distribution
$P\left(\mathcal{N}_{\eta} \big| \mathcal{N}_{\eta} \geq N c^* \right)$.
 Again, the algorithm can only explore a very small range of values of $\mathcal{N}_{\eta}$.
The difference with the previous algorithm is that 
we can now explore small regions of the form $N c^* \leq \mathcal{N}_{\eta} \leq  N c^* + \delta$
for every $c^*$, whereas we could
 before only explore the neighbourhood of the mean value
$N/2$.

The distribution of $\mathcal{N}_{\eta}$
is given by
\begin{equation}
\mathcal{P}\left(\mathcal{N}_{\eta}\right)=
\mathcal{P}\left(\mathcal{N}_{\eta} \big| \mathcal{N}_{\eta} \geq N c^* \right)
\mathcal{P}\left(\mathcal{N}_{\eta} \geq N c^* \right)\;\;\textrm{(for every $\mathcal{N}_{\eta}\geq N c^*$).}
\end{equation}
Therefore the rate function reads:
\begin{equation}
\Phi_{\eta}(c_{\eta})\equiv -\frac{\log \mathcal{P}(\mathcal{N}_{\eta}=c_{\eta} \, N)}{\beta N^2}=
 -\frac{\log \mathcal{P}(\mathcal{N}_{\eta}=c_{\eta} N \big| \mathcal{N}_{\eta} \geq N c^*)}{\beta N^2}+K_{c^*}\;\;
{\rm for }\;\; c_{\eta}>c^*
\end{equation}
where $K_{c^*}=-\frac{\log \mathcal{P}(\mathcal{N}_{\eta} \geq N c^*)}{\beta N^2}$
is a constant (independent of $c_{\eta}$).
In order to get rid of the constant $K_{c^*}$, we construct from the histogram giving $P\left(\mathcal{N}_{\eta} \big| \mathcal{N}_{\eta} \geq N c^* \right)$
 the derivative of the rate function.
This derivative is equal to $\frac{d\Phi_{\eta}(c_{\eta})}{dc_{\eta}}$. The constant $K_{c^*}$ disappears. 

We can  come back to $\Phi_{\eta}(c_{\eta})$ (and thus $P\left(\mathcal{N}_{\eta}=c_{\eta} N\right)= e^{-\beta
N^2 \Phi_{\eta}(c_{\eta})}$) from its derivative using 
an interpolation of the data for
 the derivative and a numerical integration of the interpolation.

We typically run the algorithm for $N=50$ and $10^8$ iterations.

\begin{figure}[htbp]
\includegraphics[height=6.5cm]{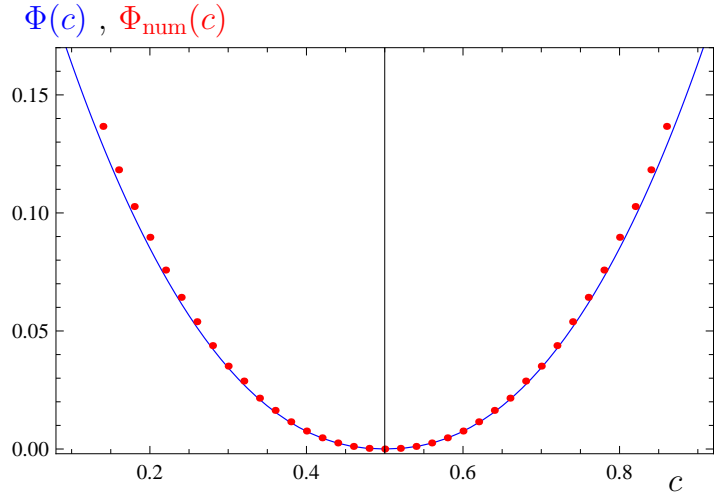}
\caption{Rate function $\Phi_{\rm num}(c)=-\log \mathcal{P}({\mathcal N}_+=cN,N)/ \left(\beta N^2\right)$ plotted as a function
of $c$ for $N=50$.
The red points are numerical data obtained with the method explained in section \ref{sec:simulnum} with $\eta=0.5$. Each point corresponds to an integer
value of $\mathcal{N}_+=c N$. The blue solid line is the analytical prediction $\Phi(c)$ given in Eq. \eqref{phic1} for the large $N$ limit.}
\label{fig:phicNum}
\end{figure}

\subsection{Back to $\mathcal{N}_+$}

Using the algorithm explained in the previous subsection, we get the distribution of $\mathcal{N}_{\eta}$
for a given value of $\eta$.
A natural way of recovering the distribution of $\mathcal{N}_+$ would be to run the algorithm for smaller and smaller values of $\eta$ as
$\mathcal{N}_+=\lim_{\eta\rightarrow 0} \mathcal{N}_{\eta}$. However, as explained above this is not an efficient method numerically.
For small $\eta$, the distribution of $\mathcal{N}_{\eta}$ is indeed not smooth; $\mathcal{N}_{+}=\mathcal{N}_{0}$ even takes integer values, i.e. it is discontinuous.

A better procedure consists in running the algorithm for a fixed (and not too small) value of $\eta$, typically
$\eta=0.5$ for $N=50$, and  exploiting the joint data that we can get for $\mathcal{N}_{\eta}$ and $\mathcal{N}_{+}$.
 When running the algorithm, we can indeed construct a joint histogram
for $\mathcal{N}_+$ and $\mathcal{N}_{\eta}$ (by keeping the value of $\mathcal{N}_+$ and $\mathcal{N}_{\eta}$ every ten steps).
With all the data for many values of the constraint $c^*$ and after having filled up the histogram by symmetry around $1/2$,
 we can then get a full histogram for $\mathcal{P}(\mathcal{N}_{+} | \mathcal{N}_{\eta} )$.

Finally we recover the distribution of $\mathcal{N}_+$ by numerical integration over $\mathcal{N}_{\eta}$:
\begin{equation}
\mathcal{P}(\mathcal{N}_{+})=\int d\mathcal{N}_{\eta}\, \mathcal{P}(\mathcal{N}_{+} | \mathcal{N}_{\eta} )  \mathcal{P}( \mathcal{N}_{\eta} )
\end{equation}

In Fig. \ref{fig:phicNum}, we plot the rate function $\Phi_{\rm num}(c)\equiv -\frac{\log \mathcal{P}({\mathcal N}_+=cN,N)}{\beta N^2}$ 
obtained numerically with the method explained above
and compare with its analytical expression $\Phi(c)$ for large $N$ given in Eq. \eqref{phic1}.
The agreement is quite good. As the distribution of  $\mathcal{N}_+$ (and $\mathcal{N}_{\eta}$ to a lesser extent) is not  smooth
for finite $N$, there are finite size effects and the convergence is a bit slow in the simulations. Therefore the agreement between numerics and the theory
is less good very far from the mean value.

\section{Probability of a gap $[\zeta_1,\zeta_2]$ in the spectrum}
\label{gap-section}

As an application of the general result derived on the two-support solution
in Section II, here we address the natural question:
what is the probability that there are no eigenvalues on the interval
$[\zeta_1,\zeta_2]$ (where $\zeta_1 \le \zeta_2$) for a Gaussian random 
matrix? As 
discussed
earlier, the natural scale for the eigenvalues of Gaussian random matrix
is $\sim \sqrt{N}$ for large $N$. Hence, it is appropriate to rescale 
$\zeta_1=w_1\sqrt{N}$ and $\zeta_2= w_2 \sqrt{N}$ and denote
this gap probability as $P(w_1,w_2,N)$ with $w_1\le w_2$.

The computation of $P(w_1,w_2,N)$ is performed in two steps.
First we fix the number of eigenvalues that are bigger than
$w_2$ to be $N_+=cN$ where $c$ denotes the fraction. Naturally
the number of eigenvalues that are less than $w_1$ is then
$N_{-}=(1-c)N$. Let $P(w_1,w_2,c,N)$ denote the gap probability
for a given fixed $c$. Then the full gap probability is obtaining
by summing over all possible values of $c$
\begin{equation}
P(w_1,w_2,N)= \int_0^1 dc\, P(w_1,w_2,c,N)
\label{gap1}
\end{equation}

The gap probability $P(w_1,w_2,c,N)$ for a fixed $c$ and for large $N$ can be 
computed exactly in the same way as the index distribution in Section II.
Once again we have the optimal charge density with two disconnected supports,
one to the left of $w_1$ and one to the right of $w_2$. Therefore, the general
solution in \eqref{fc} 
will still be valid with the only exception that in this case the edges
$l_2=w_1$ (the upper edge of the left support) and $L_1=w_2$ (the
lower edge of the right support) are already fixed. Hence
\begin{equation}
f_c^\star(x)=\frac{1}{\pi}\sqrt{\left(\frac{x-l_1}{w_1-x}\right)\,
\left( \frac{L_2-x}{w_2-x}\right)}
\:\left|\frac{w_1-L_2}{2}+\frac{w_2-l_1}{2}-x \right|
\;\;
{\rm for} \;\;  
x\in[l_1,w_1]\cup [w_2,L_2].
\label{optsol1}
\end{equation}
It remains to fix the still two unknowns 
$l_1$ (the lower edge of the left support) and $L_2$ (the upper edge
of the right support). They are fixed by the consistency
condition \eqref{constrB} which in this case reads
\begin{equation}
1+\frac{w_2^2+2 w_2 L_2 -3 L_2^2}{8}+\left(\frac{w_1-l_1}{8}\right)\left(
3 l_1+w_1+2 L_2-2 w_2 \right)=0 
\label{consis1}
\end{equation}
and the normalization condition $\int_{w_2}^{L_2} f_c^2(x) dx=c$.

One then uses this optimal solution to evaluate the saddle point
action $S[f_c^\star(x)]$ (as in \eqref{action2}) and compute the associated
large deviation function $\Phi(c,w_1,w_2)$ (which now depends
on $w_1$ and $w_2$) from \eqref{phic2}.
This gives for large $N$
\begin{equation}
P(w_1,w_2,c,N)\approx \exp\left[-\beta N^2 \Phi(c, w_1,w_2)\right].
\label{ldvc1}
\end{equation}
Substituting further this result in \eqref{gap1} and
evaluating the integral over $c$ by another saddle point one finally gets
the gap probability for large $N$
\begin{equation}
P(w_1,w_2,N)\approx \exp\left[-\beta N^2 \Psi(w_1,w_2)\right];\quad {\rm 
with}\quad \Psi(w_1,w_2)= \Phi(c^\star, w_1,w_2)
\label{gap2}
\end{equation}
where $c^\star$ minimizes the function $\Phi(c,w_1,w_2)$ over $c\in [0,1]$.
Physically the quantity $\beta N^2 \Psi(w_1,w_2)$ just represents the energy 
cost in
separating the two blobs of charges by a gap $[w_1,w_2]$ from their 
natural Wigner semicircle configuration.
 
\begin{figure}[htbp]
\includegraphics[height=8cm,width=12cm]{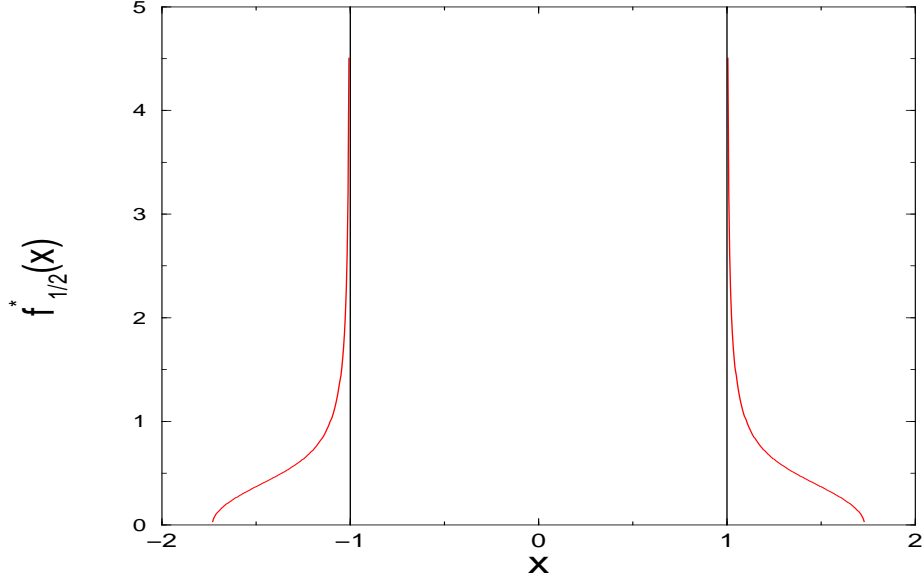}
\caption{Analytical optimal density $f_{1/2}^\star(x)$ in \eqref{optsol2}
corresponding 
to a gap over the interval $[-w,w]$ with $w=1$. The density has
two disconnected symmetrical supports over $[-\sqrt{3},-1]\cup[1,\sqrt{3}]$.
It vanishes at the upper edge $L=\sqrt{w^2+2}=\sqrt{3}$ of the right support
and at the lower edge $-L=-\sqrt{3}$ of the left support. At the edges $w_1=-1$
and $w_2=1$, the density has an inverse square root divergence.}
\label{fig:gapden}
\end{figure}

In principle one can compute the large deviation function $\Psi(w_1,w_2)$
for arbitrary $[w_1,w_2]$ by following the above procedure. Here, for 
simplicity, we present the explicit result for the simple case when
the two walls are placed symmetrically around the origin: $w_1=-w$ and 
$w_2=w$. In this case, it is evident due to the symmetry that the 
optimal value must be $c^\star=1/2$. The optimal solution in 
\eqref{optsol1} for $c=1/2$ is also symmetric around $x=0$
with $l_1=-L$ and $L_2=L$ and has the simple form
\begin{equation}
f_{1/2}^\star(x)= \frac{1}{\pi}\sqrt{\frac{L^2-x^2}{x^2-w^2}}\, |x|\;\;
{\rm for} \;\;  
x\in[-L,-w]\cup [w,L]
\label{optsol2}
\end{equation}
The only unknown $L$ is fixed by the normalization condition
$\int_w^L f_{1/2}^\star(x) dx =1/2$. This uniquely fixes 
\begin{equation}
L=\sqrt{w^2+2}. 
\label{lw}
\end{equation}
A plot of this solution is provided in Fig. (\ref{fig:gapden}). Note that
when $w\to 0$, $f_{1/2}^\star(x)= \sqrt{2-x^2}/\pi$ reduces to the Wigner
semicircle as one would expect, because without any constraint 
the semicircle form is the natural optimal density for $c=1/2$.

Having determined the optimal solution explicitly, we next proceed to compute
the large deviation function $\Phi(c,-w,w)$ from \eqref{phic2}. For this we 
need to evaluate the second moment $\mu_2$
and the two Lagrange multipliers $A_1$ and $A_2$. Using 
\eqref{optsol2} one can easily evaluate the second moment
\begin{equation}
\mu_2= \int_{-\infty}^{\infty} x^2 f_{1/2}^\star(x) dx= w^2+\frac{1}{2}.
\label{mu2}
\end{equation}
To fix the Lagrange multipliers, we substitute $x=L$ and $x=-L$ 
in \eqref{inteq1} to get two equations
\begin{align}
L^2+A_1+A_2&= 2 \int_{-\infty}^{\infty} f_{1/2}^\star(x') \log(L-x')\,dx' \\
L^2+A_2&=  2 \int_{-\infty}^{\infty} f_{1/2}^\star(x') \log(L+x')\,dx'
\label{lagrange1}
\end{align}
Using the explicit form of $f_{1/2}^\star(x)$ in \eqref{optsol2} it is easy to 
verify that both integrals on the right hand side are identical and are given 
by
\begin{equation}
I=\int_{-\infty}^{\infty} f_{1/2}^\star(x')\, \log(L-x')\,dx'=\int_{w}^L  
f_{1/2}^\star(x')\,\log(L^2-x^2)\, dx'=\frac{1-\log 2}{2}.
\label{I1}
\end{equation}
Solving these two linear equations, we get
\begin{equation}
A_1=0; \quad {\rm and}\quad A_2= 2I-L^2=-1-\log 2-w^2.
\label{lagrange2}
\end{equation}
Substituting the values of $\mu_2$, $A_1$ and $A_2$ in \eqref{phic2} gives
a very simple expression
\begin{equation}
\Phi(c=1/2,-w,w)= \frac{w^2}{2};\quad {\rm hence }\quad 
\Psi(-w,w)=\frac{w^2}{2}.
\label{Psiw}
\end{equation}
This leads to the result that the probability that there are no
eigenvalues in the interval $[-w,w]$ for a Gaussian random matrix
in the limit of large $N$ 
is simply
\begin{equation}
P(-w,w,N)\approx \exp\left[-\frac{\beta}{2}w^2 N^2\right].
\label{gap3}
\end{equation}
Note that when $w\to 0$, the probability approaches to $1$ which is to
be expected since without any constraint the system naturally settles
into the Wigner semicircle which is gapless at the origin.

\section{A formula for the variance of the index at finite $N$ for $\beta=2$}
\label{variance-section}

So far, we have computed the index distribution in the large $N$ limit.
From this result, we were able to show that the variance of the 
number of positive eigenvalues
\begin{equation}\label{stdsigma}
\Delta(N)=\langle (\mathcal{N}_{+}-N/2)^2\rangle
\end{equation}
increases logarithmically with $N$ to leading order for large $N$ as
in \eqref{variance}. A natural question is if one can derive 
an exact formula for the variance for {\em finite} $N$ and not just
for large $N$. In this section, we show that at least in the special
case $\beta=2$, it is possible   
to derive an exact formula for the variance valid at
fixed and finite $N$ and is given by
\begin{equation}\label{formula variance finite N}
\Delta(N)=\frac{\mathbf{Z}_N^{\prime\prime}(0)}{\mathbf{Z}_N(0)}-\frac{N^2}{4}
\end{equation}
where:
\begin{equation}\label{defZNP}
\mathbf{Z}_N(p)=\det\left[\left(e^{-p}+(-1)^{i+j}\right)\Gamma\left(\frac{i+j-1}{2}\right)\right]_{i,j=1,\ldots,N}
\end{equation}
and $(.)^\prime$ denotes differentiation with respect to $p$.

In order to prove \eqref{formula variance finite N}, we start from the pdf \eqref{probdef} :
\begin{equation}
\label{probdefagain}
\mathcal{P}(\mathcal{N}_{+},N)=\frac{1}{Z_N}\int_{(-\infty,\infty)^N}\prod_i d\lambda_i e^{-\frac{\beta}{2}\sum_{i=1}^N\lambda_i^2}\prod_{j<k}|\lambda_j-\lambda_k|^\beta
\delta\left(\mathcal{N}_{+}-\sum_{i=1}^N \theta(\lambda_i)\right)
\end{equation}

and define its moment generating function (Laplace transform) as:
\begin{equation}\label{Znp}
\mathcal{Z}_N(p)=\int_{(-\infty,\infty)^N}\prod_{i=1}^N d\lambda_i e^{-\sum_{i=1}^N\lambda_i^2-p\sum_{i=1}^N\theta(\lambda_i)}\prod_{j<k}|\lambda_j-\lambda_k|^2
\end{equation}

We are going to prove that:
\begin{equation}\label{prove}
\mathcal{Z}_N(p)=\frac{N!}{2^N}\mathbf{Z}_N(p)
\end{equation}
where $\mathbf{Z}_N(p)$ is given in \eqref{defZNP}.
On the other hand, it is easy to see that:
\begin{equation}\label{fins}
\Delta(N)=\frac{\mathcal{Z}_N^{\prime\prime}(0)}{\mathcal{Z}_N(0)}-\frac{N^2}{4}
\end{equation}
Combining \eqref{prove} with \eqref{fins} we readily obtain \eqref{formula variance finite N}.

In order to prove \eqref{prove}, we start from \eqref{Znp}:
\begin{equation}\label{Znp2}
\mathcal{Z}_N(p)=\int_{(-\infty,\infty)^N}\prod_{i=1}^N d\lambda_i e^{-\sum_{i=1}^N\lambda_i^2-p\sum_{i=1}^N\theta(\lambda_i)}\prod_{j<k}|\lambda_j-\lambda_k|^2
\end{equation}
We can write the square of the Vandermonde determinant in \eqref{Znp2} as:
\begin{equation}
\prod_{j<k}|\lambda_j-\lambda_k|^2 = \det(A_k(\lambda_j))\det(B_k(\lambda_j))
\end{equation}
with $A_k(x)=B_k(x)=x^{k-1}$, and then apply the Andr\'eief identity \cite{and}:
\begin{equation}
\int\prod_{i=1}^N d\mu(\lambda_i)\det(A_k(\lambda_j))\det(B_k(\lambda_j))=N!\det\left(\int d\mu(x)A_k(x)B_j(x)\right)
\end{equation} 
valid for a benign integration measure $\mu(x)$. In our case, we have $\mu(x)=e^{-x^2-p\theta(x)}$, leading to:
\begin{equation}
\mathcal{Z}_N(p)=N!\det\left(\int_{-\infty}^\infty dx\ e^{-x^2-p\theta(x)} x^{k+j-2}\right)
\end{equation}
Evaluating the integral, we get immediately to Eq. \eqref{prove}.

From this determinantal representation for the variance in Eq. \eqref{formula variance finite N}, 
Prellberg~\cite{Prellberg} noted 
that the following exact formula for the variance of the index for $\beta=2$ holds:
 \begin{equation}\label{main}
\Delta(N)=\frac{N}{4}-\frac{2}{\pi^2}\sum_{\stackrel{1\leq i<j\leq N}{i+j\mbox{ odd}}}\frac{1}{(i-j)^2}\frac{\Gamma\left(\lfloor\frac{i}{2}\rfloor
+\frac{1}{2}\right)\Gamma\left(\lfloor\frac{j}{2}\rfloor
+\frac{1}{2}\right)}{\Gamma\left(\lfloor\frac{i+1}{2}\rfloor
\right)\Gamma\left(\lfloor\frac{j+1}{2}\rfloor\right)}
\end{equation}
where $\lfloor x\rfloor$ stands for the greatest integer less or equal to $x$, and $\Gamma(x)$ is the Gamma function.

It is convenient to group the terms in the sum for even and odd terms. In this way we can perform one of the sums and we can write for even $N$ \eqref{main} as:
\begin{equation}\label{deltasum}
\Delta(N)=\frac{N}{4}-\frac{2}{\pi^2}\sum_{m=0}^{N/2-1}t_m
\end{equation}
where:
\begin{equation}
t_m=\frac{\Gamma(m+1/2)^2}{\Gamma(m)\Gamma(m+1)} \ _4F_3\left(\frac{1}{2},\frac{1}{2},1,1-m;\frac{3}{2},\frac{3}{2},\frac{1}{2}-m|1\right)+\frac{\Gamma(m+1/2)\Gamma(m+3/2)}{\Gamma(m+1)^2} \ _4F_3\left(\frac{1}{2},\frac{1}{2},1,-m;\frac{3}{2},\frac{3}{2},\frac{1}{2}-m|1\right).
\end{equation}
where $_4F_3$ is a generalized hypergeometric function. As this expression is complicated to the point of being useless (except for numerical analyses) we look for an integral representation for $t_m$. We achieve this by writing the defining series expansion for the hypergeometric function, using an integral representation for the gamma functions in its coefficients and then exchanging the integral and the sum. The final result is expressed as an integral over a new variable $t\in[0,1]$ as:  
\begin{equation}
\label{tm1}
t_m=\frac{\pi^2}{2}-\frac{1}{2\sqrt{\pi}}\frac{(m-1/2)!}{m!} \underbrace{\int_0^1 dt\ \frac{t^{m}}{\sqrt{1-t}}[\tanh^{-1}(\sqrt{t})+(2m+1)(\mathrm{Li}_2 (\sqrt{t})-\mathrm{Li}_2(-\sqrt{t}))]}_{\mathcal{I}_m},
\end{equation}
where $\mathrm{Li}_2 (z)=\sum_{k=0}^\infty\frac{z^k}{k^2}$ is the Polylogarithm function.

\begin{figure}[htbp]
\includegraphics[height=6cm]{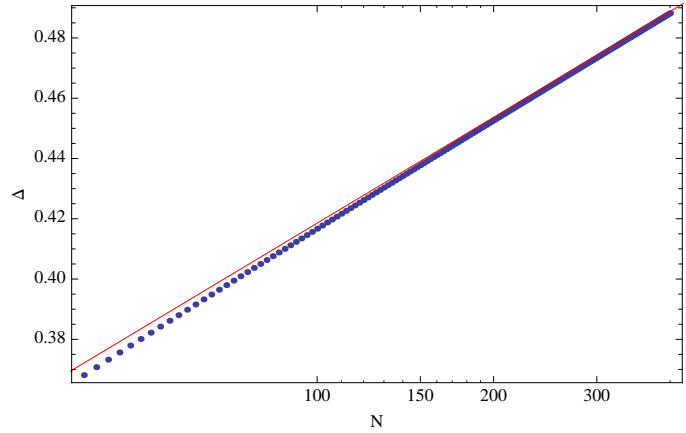}
\caption{The variance of the index $\Delta(N)$ as a function of
$\log(N)$ for $\beta=2$ (dotted, exact finite $N$ formula in \eqref{deltasum}; 
solid, large $N$ in \eqref{fullasym}).
A linear fit for the former gives $\Delta(N)\simeq 0.052
\log N + 0.184 $. The prefactor $0.052$ is in good agreement with the 
leading
theoretical
prefactor $(2\pi^2)^{-1}\simeq 0.051$, and the 
constant correction
term $0.184$ is also in good agreement with the theoretical constant $C$ in \eqref{Ctheo}.}
\label{fig:variance}
\end{figure}

Now, we separate the integral $\mathcal{I}_m$ in two terms,
\begin{equation}
\mathcal{I}_m=\mathcal{I}^{(1)}_m+\mathcal{I}^{(2)}_m=\int_0^1 dt\frac{t^{m}}{\sqrt{1-t}}\tanh^{-1}(\sqrt{t})+\int_0^1 dt \frac{t^m}{\sqrt{1-t}}(2m+1)\left[\mathrm{Li}_2(\sqrt{t})-\mathrm{Li}_2(-\sqrt{t})\right]
\end{equation}
and we separate further the second integral as
\begin{align}
\nonumber\mathcal{I}^{(2)}_m &=\int_0^1 dt \frac{t^m}{\sqrt{1-t}}(2m+1)\left[\mathrm{Li}_2(\sqrt{t})-\mathrm{Li}_2(-\sqrt{t})-\pi^2/4\right]+\frac{\pi^2}{4}(2m+1)\int_0^1dt\frac{t^m}{\sqrt{1-t}}\\
&=\int_0^1 dt \frac{t^m}{\sqrt{1-t}}(2m+1)\left[\mathrm{Li}_2(\sqrt{t})-\mathrm{Li}_2(-\sqrt{t})-\pi^2/4\right]+\frac{\pi^{5/2}m!}{2(m-1/2)!}.
\end{align}
The last term, when inserted back in \eqref{tm1}, cancels half of the constant $\pi^2/2$ in $t_m$ so we are left with
\begin{equation}
t_m=\frac{\pi^2}{4}-\frac{(m-1/2)!}{2\pi^{1/2}m!}\int_0^1dt \frac{t^m}{\sqrt{1-t}}\left\{\tanh^{-1}(\sqrt{t})+(2m+1)\left[\mathrm{Li}_2(\sqrt{t})-\mathrm{Li}_2(-\sqrt{t})-\frac{\pi^2}{4}\right]\right\}.
\end{equation}
An integration by parts on the term in the integrand linear in $m$ (considering that $\partial_t\left[\mathrm{Li}_2(\sqrt{t})-\mathrm{Li}_2(-\sqrt{t})\right]=\tanh^{-1}(\sqrt{t})/t$) gives:
\begin{equation}
t_m=\frac{\pi^2}{4}-\frac{(m-1/2)!}{2\pi^{1/2}m!}\int_0^1dt \frac{t^m}{\sqrt{1-t}}\left[\frac{1}{1-t}\left(\frac{\pi^2}{4}+{\rm Li}_2(-\sqrt{t})-{\rm Li}_2(\sqrt{t})\right)-\tanh^{-1}(\sqrt{t})\right].
\end{equation}
We need to sum this expression over $m=0,...,N/2-1$ to get $\Delta(N)$. The constant term in $t_m$ will cancel against the linear term $N/4$ in $\Delta$ and a compact integral representation for $\Delta(N)$ can now be obtained exchanging the order of integration over $t$ and summation over $m$:
\begin{equation}
\Delta(N)=\frac{1}{\pi^{5/2}}\int_0^1dt\ K(t,N)\frac{1}{\sqrt{1-t}}\left[\frac{1}{1-t}\left(\frac{\pi^2}{4}+{\rm Li}_2(-\sqrt{t})-{\rm Li}_2(\sqrt{t})\right)-\tanh^{-1}(\sqrt{t})\right],
\end{equation}
where
\begin{equation}
K(t,N)=\sum_{m=0}^{N/2-1}\frac{(m-1/2)!}{m!}t^m=(1-t)^{-1/2}\left(\sqrt{\pi}-\mathrm{B}(t;N/2,1/2)\frac{\Gamma\left(\frac{1+N}{2}\right)}{\Gamma(N/2)}\right),
\end{equation}
where $\mathrm{B}$ is the incomplete Euler beta function, defined as $\mathrm{B}(z;a,b)=\int_0^z d\tau\ \tau^{a-1}(1-\tau)^{b-1}$.

This representation turns out to be very useful to pull out the large $N$ logarithmic growth of $\Delta (N)$ and the constant term (and possibly could yield a complete asymptotic expansion in $1/N$). In order to do this we notice that for large $N$ the function $K(t,N)$ is concentrated near $t=1$. So we expand the remaining integrand to lowest order in $1-t$ obtaining the leading order and part of the constant term as:
\begin{equation}
\frac{1}{2\pi^{5/2}}\int_0^1 dt\ K(t,N)(1-t)^{-1/2}=\frac{1}{2\pi^2}\log N+\frac{1}{2\pi^2}(\gamma+\log 2)+\mathcal{O}(N^{-1}).
\end{equation}
where $\gamma=0.577215...$ is Euler's constant.

One can prove that the remaining terms in the expansion in powers of $(1-t)$ contribute to $\mathcal{O}(1)$ but not to the leading logarithm. We can formally lump these terms together and we can write the
asymptotic law for $\Delta(N)$ as:
\begin{equation}\label{fullasym}
\Delta(N)=\frac{1}{2\pi^2}\log N+C+\mathcal{O}(N^{-1}),
\end{equation}
where the constant $C$ is:
\begin{equation}
C=\frac{1}{2\pi^2}(\gamma+\log 2)+\lim_{N\to\infty}\frac{1}{\pi^{5/2}}\int_0^1 dt\ K(t,N)\frac{1}{\sqrt{1-t}}\left[\frac{1}{1-t}\left(\frac{\pi^2}{4}+{\rm Li}_2(-\sqrt{t})-{\rm Li}_2(\sqrt{t})\right)-\tanh^{-1}(\sqrt{t})-\frac{1}{2}\right].
\end{equation}
Now the limit $N\to\infty$ can be taken safely inside the integral (Euler's $\mathrm{B}$ function goes to zero) and we are left with the following nontrivial constant:
\begin{eqnarray}\label{Ctheo}
\nonumber C&=&\frac{1}{2\pi^2}(\gamma+\log 2)+\frac{1}{\pi^2}\int_0^1dt \frac{-2+\pi^2+2t-4(1-t)\tanh^{-1}(\sqrt{t})-4{\rm Li}_2(\sqrt{t})+4{\rm Li}_2(-\sqrt{t})}{4(1-t)^2}\\
&=& \frac{\gamma+1+3\log 2}{2\pi^2}=0.1852484182... \ .
\end{eqnarray}
where in the last step we have performed one extra integration by part. The constant $C$ is in good agreement with the fit of the finite $N$ results for large $N$ (see fig. \ref{fig:variance}). A careful series expansion of $K(t,N)$ for large $N$ should give the complete $1/N$ expansion of $\Delta(N)$. This is left for future work.

\section{Conclusions} In summary, we have computed for large $N$ the 
probability that a Gaussian matrix $N\times N$ with real spectrum has a 
fraction $c$ of positive eigenvalues. Using a Coulomb gas method, a large 
deviation principle for this probability can be formulated. In physical 
terms, the problem amounts to finding the free energy of a system of charged 
particles repelling each other via a 2d Coulomb interaction and confined into 
a quadratic well, with the constraint that a fraction $c$ of them is kept on 
the positive semiaxis. Due to the long-range nature of the interaction, the 
free energy is super-extensive in the number of particles, and scales as 
$\sim\mathcal{O}(N^2)$, as it is customary in this type of problems.  
We have computed explicitly the large deviation function $\Phi(c)$, which 
quantifies the 
rate 
of occurrence of unusual fluctuations of the index, 
for all $0\le c\le 1$. This function has a minimum at $c=1/2$ around 
which it has a quadratic form modulated by a logarithmic singularity.
This logarithmic singularity leads to the result that  
the variance of the index displays a logarithmic growth with the matrix size 
$N$ for all $\beta$. For $\beta=2$, we have found a representation of the variance in terms of derivatives of a certain Hankel determinant.
Based on this representation, Prellberg~\cite{Prellberg} was able to give an explicit expression 
for the index variance involving a finite double sum. We performed an asymptotic analysis of Prellberg's finite $N$ expression,
whose leading behavior is precisely $\sim (2\pi^2)^{-1}\log(N)$, in perfect agreement with our Coulomb gas result. In addition,
we determined exactly the constant term $C$ in the expansion, which turns out to be a highly non-trivial value as in Eq. \eqref{Ctheo}.

We have also presented a general method to obtain explicitly a two-support
solution of a singular integral equation of the form \eqref{intequation1}. This 
method
consists in iterating the single support Tricomi solution twice. We have 
demonstrated how this method can be used to compute the probability
of a gap $[\zeta_1, \zeta_2]$ in the spectrum of the eigenvalues. Given
the fact that singular integral equation of the type \eqref{intequation1}
occurs quite generically for other random matrices (such as Wishart
matrices~\cite{Wishart}), we expect that this method will be useful
in a broad variety of applications. 

\begin{acknowledgments}
We thank T. Prellberg for sharing with us his results about the explicit evaluation of $\Delta(N)$ in terms of a double sum.
\end{acknowledgments}


\begin{thebibliography}{99}

\bibitem{Wigner} E.P. Wigner, Proc. Cambridge Philos. Soc. {\bf 47}, 790 
(1951).

\bibitem{Mehta} M.L. Mehta, {\it Random Matrices} (Academic Press, Boston, 1991).

\bibitem{DM} D.S. Dean and S.N. Majumdar, Phys. Rev. Lett. {\bf 97}, 160201 
(2006); Phys. Rev. E {\bf 77}, 41108 (2008).

\bibitem{majumdar:060601} S.N. Majumdar and M. Vergassola, Phys. Rev. Lett. 
{\bf 102}, 060601 (2009).

\bibitem{BEMN} G. Borot, B. Eynard, S.N. Majumdar, and C. Nadal, {\em Preprint} [arXiv: 
1009.1945] (2010).

\bibitem{Saito} N. Saito, Y. Iba, and K. Hukushima, Phys. Rev. E {\bf 82}, 
031142 (2010).

\bibitem{ben} G. Ben Arous, A. Dembo, and A. Guionnet, Prob. Theory Related Fields, {\bf 2}, 73 (2001).

\bibitem{vivo2007large} P. Vivo, S.N. Majumdar, and O. Bohigas, J. Phys. A: 
Math. Theor. {\bf 40}, 4317 (2007).

\bibitem{KC} E. Katzav and I.P. Castillo, Phys. Rev. E {\bf 82}, 041004 
(2010). 

\bibitem{BrayDean} A.J. Bray and D.S. Dean, Phys. Rev. Lett. {\bf 98}, 150201 
(2007).

\bibitem{fyod} Y.V. Fyodorov and I. Williams, J. Stat. Phys. {\bf 129}, 1081 
(2007).

\bibitem{PR} G. Parisi and T. Rizzo, Phys. Rev. Lett. {\bf 101}, 117205 
(2008); Phys. Rev. B {\bf 79}, 134205 (2009). 

\bibitem{MG} C. Monthus and T. Garel, J. Stat. Mech.: Th. and Exp. {\bf 
P02023} (2010).


\bibitem{vivo:216809} P. Vivo, S.N. Majumdar, and O. Bohigas, 
Phys. Rev. Lett. {\bf 101}, 216809 (2008).

\bibitem{conductance} P. Vivo, S.N. Majumdar, and O. Bohigas,
Phys. Rev. B. {\bf 81}, 104202 (2010).


\bibitem{facchi:050502} P. Facchi, U. Marzolino, G. Parisi, 
S. Pascazio, and A. Scardicchio, Phys. Rev. Lett. {\bf 101}, 050502 (2008).

\bibitem{NMV} C. Nadal, S.N. Majumdar, and M. Vergassola, Phys. Rev. Lett., 
{\bf 104}, 110501 (2010); see also {\em Preprint} [arXiv:1006.4091] (2010).

\bibitem{pasquale} A. De Pasquale, P. Facchi, G. Parisi, S. Pascazio,
and A. Scardicchio, Phys. Rev. A {\bf 81}, 052324 (2009).

\bibitem{pp2010} P. Vivo, {\em Preprint} [arXiv:1009.1517] (2010).

\bibitem{kaz} P. Kazakopoulos, P. Mertikopoulos, A.L. Moustakas and G. Caire, 
{\em Preprint} [arXiv:0907.5024] (2009).

\bibitem{Schehr1} G. Schehr, S.N. Majumdar, A. Comtet, and J. Randon-Furling,
Phys. Rev. Lett. {\bf 101}, 150601 (2008).

\bibitem{NM} C. Nadal and S.N. Majumdar, Phys. Rev. E, {\bf 79}, 
061117 (2009).

\bibitem{Schehr2} J. Rambeau and G. Schehr, Europhys. Lett. {\bf 91}, 60006 
(2010).

\bibitem{FMS} P.J. Forrester, S.N. Majumdar and G. Schehr, {\em Preprint} [arXiv:1009.2362] (2010).

\bibitem{letter} S.N. Majumdar, C. Nadal, 
A. Scardicchio, and P. Vivo, Phys. Rev. Lett. {\bf 103}, 220603 (2009).


\bibitem{Dyson} F.J. Dyson, J. Math. Phys. {\bf 3}, 140 (1962); {\bf 3}, 157 
(1962); {\bf 3}, 166 (1962).

\bibitem{Dyson-Mehta} F.J. Dyson and M.L. Mehta, J. Math. Phys. {\bf 3}, 701 
(1962).

\bibitem{CL} O. Costin and J.L. Lebowitz, Phys. Rev. Lett. {\bf 75}, 69 
(1995).

\bibitem{FS} M.M. Fogler and B.I. Shklovskii, Phys. Rev. Lett. {\bf 74}, 3312 
(1995).


\bibitem{Wales} D.J. Wales, {\it Energy Landscapes: 
Applications to Clusters, Biomolecules and Glasses} (Cambridge University 
Press, 2004).

\bibitem{CGG} A. Cavagna, J.P. Garrahan, and I. Giardina, Phys. Rev. B {\bf 61}, 
3960 (2000).

\bibitem{Douglas} M.R. Douglas, JHEP {\bf 05}, 046 (2003).

\bibitem{aazami} A. Aazami and R. Easther, JCAP03 p. 013 (2006).

\bibitem{Mersini-Houghton05}
L. Mersini-Houghton, Class. Quant. Grav. {\bf 22},  3481  (2005).

\bibitem{randpol} J.-P. Dedieu and G. Malajovich, Journal of Complexity {\bf 24}, 89 (2008).



\bibitem{Forrester} P.J. Forrester, {\em Log-gases and random matrices} 
(Princeton University Press, Princeton, NJ, 2010).



\bibitem{Tricomi} F.G. Tricomi, {\it Integral Equations} (Pure Appl. Math V, Interscience, London, 1957).


\bibitem{BIPZ} E. Brezin, C. Itzykson, G. Parisi, and J.B. Zuber, Commun.
Math. Phys. {\bf 59}, 35 (1978).

\bibitem{akemann} G. Akemann, Nucl. Phys. B {\bf 507}, 475 (1997). 

\bibitem{and} C. Andr\'eief, M\'em. de la Soc. Sci., Bordeaux {\bf 2}, 1 
(1883). 


\bibitem{Prellberg} T. Prellberg, private communication.

\bibitem{Wishart} J. Wishart, Biometrika {\bf 20}, 32 (1928). 













\end{thebibliography}
\end{document}